\documentclass[prb,aps,amssymb,twocolumn,superscriptaddress,notitlepage]{revtex4-1}
\usepackage{amsmath}
\usepackage{amssymb}
\usepackage{amsthm}
\usepackage{amsfonts}
\usepackage{listings}
\usepackage{enumerate}
\usepackage{latexsym}
\usepackage{psfrag}
\usepackage{bm}
\usepackage{graphicx}
\usepackage[caption=false]{subfig}
\usepackage{blkarray}
\usepackage{array}
\usepackage{color}
\usepackage[normalem]{ulem}
\usepackage{hyperref}

\def\Gq{G_{\mathbf{q}}}
\def \Gqp{G_{\mathbf{q}'}}

\makeatletter
\DeclareRobustCommand{\element}[1]{\@element#1\@nil}
\def\@element#1#2\@nil{%
  #1%
  \if\relax#2\relax\else\MakeLowercase{#2}\fi}
\makeatother

\newcolumntype{L}{>{$}l<{$}}

\newtheorem{defn}{Definition}
\newtheorem*{defn*}{Definition}
\newtheorem{conditions}{}
\newtheorem{prop}{Proposition}
\newtheorem{cor}{Corollary}

\newcommand{\half}{\frac{1}{2}}

\begin{document}
\title{Building Blocks of Topological Quantum Chemistry:\\ Elementary Band Representations}
\author{Jennifer Cano}
\thanks{These authors contributed equally to the preparation of this work.}
\affiliation{Princeton Center for Theoretical Science, Princeton University, Princeton, New Jersey 08544, USA}
\author{Barry Bradlyn}
\thanks{These authors contributed equally to the preparation of this work.}
\affiliation{Princeton Center for Theoretical Science, Princeton University, Princeton, New Jersey 08544, USA}
\author{Zhijun Wang}
\thanks{These authors contributed equally to the preparation of this work.}
\affiliation{Department of Physics, Princeton University, Princeton, New Jersey 08544, USA}
\author{L. Elcoro}
\affiliation{Department of Condensed Matter Physics, University of the Basque Country UPV/EHU, Apartado 644, 48080 Bilbao, Spain}
\author{M.~G. Vergniory}
\affiliation{Donostia International Physics Center, P. Manuel de Lardizabal 4, 20018 Donostia-San Sebasti\'{a}n, Spain}
\affiliation{Department of Applied Physics II, University of the Basque Country UPV/EHU, Apartado 644, 48080 Bilbao, Spain}
\affiliation{Ikerbasque, Basque Foundation for Science, 48013 Bilbao, Spain}
\author{C. Felser}
\affiliation{Max Planck Institute for Chemical Physics of Solids, 01187 Dresden, Germany}
\author{M.~I.~Aroyo}
\affiliation{Department of Condensed Matter Physics, University of the Basque Country UPV/EHU, Apartado 644, 48080 Bilbao, Spain}
\author{B. Andrei Bernevig}
\thanks{Permanent Address: Department of Physics, Princeton University, Princeton, New Jersey 08544, USA }
\affiliation{Department of Physics, Princeton University, Princeton, New Jersey 08544, USA}
\affiliation{Donostia International Physics Center, P. Manuel de Lardizabal 4, 20018 Donostia-San Sebasti\'{a}n, Spain}
\affiliation{Laboratoire Pierre Aigrain, Ecole Normale Sup\'{e}rieure-PSL Research University, CNRS, Universit\'{e} Pierre et Marie Curie-Sorbonne Universit\'{e}s, Universit\'{e} Paris Diderot-Sorbonne Paris Cit\'{e}, 24 rue Lhomond, 75231 Paris Cedex 05, France}
\affiliation{Sorbonne Universit\'{e}s, UPMC Univ Paris 06, UMR 7589, LPTHE, F-75005, Paris, France}
\affiliation{LPTMS, CNRS (UMR 8626), Universit\'e Paris-Saclay, 15 rue Georges Cl\'emenceau,\\ 91405 Orsay, France}
\date{\today}
\begin{abstract}
The link between chemical orbitals described by local degrees of freedom and band theory, which is defined in momentum space, was proposed by Zak several decades ago for spinless systems with and without time-reversal in his theory of ``elementary'' band representations. 
In a recent paper [Bradlyn et al., Nature {\bf 547}, 298--305 (2017)], we introduced the generalization of this theory to the experimentally relevant situation of spin-orbit coupled systems with time-reversal symmetry and proved that all bands that do not transform as band representations are topological. 
Here, we give the full details of this construction.
We prove that elementary band representations are either connected as bands in the Brillouin zone and are described by localized Wannier orbitals respecting the symmetries of the lattice (including time-reversal when applicable), or, if disconnected, describe topological insulators. 
We then show how to generate a band representation from a particular Wyckoff position and determine which Wyckoff positions generate elementary band representations for all space groups.
This theory applies to spinful and spinless systems, in all dimensions, with and without time reversal. 
We introduce a homotopic notion of equivalence and show that it results in a finer classification of topological phases than approaches based only on the symmetry of wavefunctions at special points in the Brillouin zone.
Utilizing a mapping of the band connectivity into a graph theory problem, we show in companion papers which Wyckoff positions can generate disconnected elementary band representations, furnishing a natural avenue for a systematic materials search.
\end{abstract}
\maketitle

\section{Introduction}

Stoichometric crystalline materials in nature consist of ordered arrays of atoms at lattice sites, with electrons in local orbitals that hybridize and largely determine many physical properties of the material. Because of the non-vanishing overlap of orbitals, the real-space Hamiltonian of a crystal contains terms coupling different lattice sites. Hence, while being local, a crystal Hamiltonian is not diagonal in real space. Though the chemical description and many physical properties are local, physicists have chosen to understand crystals using band theory because the Hamiltonian and its associated Schr\"{o}dinger equation are diagonal in momentum space. The momentum space picture, while extremely useful, also obfuscates the local physics present in crystals.

To remedy this disconnect, Zak\cite{Zak1980,Zak1981} introduced the concept of a band representation (BR) for spinless systems, with and without time-reversal symmetry. These band representations are, roughly, mathematical vehicles that relate the orbital representation of the electrons on sites in real (direct) space to the momentum space description of the electron bands in the Brillouin zone. Zak realized that band representations can be decomposed into what he called ``elementary building bricks,''\cite{Zak1980} which are themselves band representations, but which cannot be further subdivided while preserving the symmetry operations of the system. Twenty years later, in a series of papers, Zak and Michel examined the connectivity of these Elementary Band Representations (EBRs) for spinless systems.\cite{Michel1999,Michel2000,Michel2001} Physically, the EBR connectivity represents the number of energy bands that are connected together in the Brillouin Zone (BZ), and which cannot be disconnected without breaking the space-group symmetry of the crystal.

While Zak and Michel sought to prove that EBRs were connected, in this manuscript we show that disconnected EBRs do exist. A disconnected EBR consists of disconnected bands that together form the EBR.
We have previously introduced these concepts as part of a much broader paper\cite{NaturePaper}.
In that paper, we argued that several ingredients, one of them being the EBRs, can be used to define a new field, Topological Quantum Chemistry, which offers an unprecedented understanding and predictive power about topological materials.
In the current paper we extend and fill in all the necessary proofs in the theory of band representations. 
Our main result is that topological bands are exactly bands that do not form band representations. 
Among these, a particularly important set of topological bands are those that form one connected component {(subpart)} of a disconnected EBR.  These bands \emph{must always} be topological.
Such connected groups of bands that cannot result from a localized set of atomic orbitals that obey the crystal symmetry -- i.e., which lack an ``atomic limit'' -- were studied in a related context in Refs~\onlinecite{Parameswaran2012,Po2016}.

The study of systems with spin-dependent terms in the Hamiltonian require double space groups and their single-and double-valued irreps.\cite{GroupTheoryPaper,Cracknell}
In this manuscript, when we refer to space groups, we are implicitly referring to double space groups, unless specified otherwise.
We generalize Zak's theory to double-valued group representations with and without time-reversal (TR) symmetry.
We also derive necessary and sufficient conditions for band representations to be elementary, and highlight certain ``exceptional cases.'' We tabulate these exceptions for spinful systems with and without TR and fill in some cases missed in the original table of exceptions provided by Michel and Zak for spinless systems.
With this information, we can find the minimum set of Wyckoff positions necessary to generate all EBRs, {for single and double groups, with and without TR,} an enormous task that we tabulate in related manuscripts.\cite{GraphTheoryPaper,GraphDataPaper}.  When the EBRs are connected, we show that they are related to exponentially localized Wannier orbitals that respect the symmetries of the crystal (plus, when applicable, time-reversal).  When they are fractionally filled, connected EBRs represent protected  (semi-)metals. Our theory, along with the tables presented in Ref~\onlinecite{GraphDataPaper}, gives a full analysis of the possible fillings of protected metals that exist in nature. 
By charge transfer, fillings such as 1/16, 1/8, and 1/24 exist, the last being the lowest possible filling. When a connected EBR is fully filled, it represents a group of topologically trivial bands, which describe an atomic limit. As a corollary, we show there exist different atomic limits, not adiabatically continuable to each other, but all described by symmetric, exponentially localized Wannier states; they can be differentiated by the value of their Berry phases\cite{Bacry1988b,Bacry1993,Michel1992,Holler2017} (Wilson loops\cite{Yu11,Soluyanov2011,Alexandradinata16,ArisCohomology}) or Berry phases of Berry phases,\cite{hughesbernevig2016} etc.
When an EBR is disconnected, i.e. when it is formed by bands separated from each other by an energy gap, then we show that at least one of those groups of bands must be topological.

Our theory provides a crucial first step in an informed search for topological materials: namely, searching for materials whose orbitals, especially at the Fermi level, induce disconnected EBRs.
To this end, in a series of related works we have defined, categorized and given the representation data for all the possible EBRs in all space groups: single- and double-valued representations (which describe systems with and without spin-orbit coupling (SOC)) and with and without TR symmetry.\cite{NaturePaper,GroupTheoryPaper,GraphDataPaper,GraphTheoryPaper}
We find that there are 10,403 of these different EBRs {(3,383 single-valued and 2,263 double-valued without TR and 3,141 single-valued and 1,616 double-valued with TR imposed).
The EBRs and their irreps at high-symmetry points are freely available on the Bilbao Crystallographic Server.\cite{GroupTheoryPaper}}

We now describe the connection to previous works classifying topological crystalline materials.
Starting with the inversion eigenvalue characterization of topological insulators\cite{Fu2007}, eigenvalues of crystal symmetry operations have been used to characterize $\mathbb{Z}_2$ topological insulators\cite{Teo08} and to compute the Chern number of a set of bands.\cite{Hughes11,Fang2012}
It was later shown that crystal symmetries can enhance the Altland-Zirnbauer classification of topological insulators\cite{Teo08,Turner2010,Fu2011,Turner2012,Morimoto2013,Chiu2013,Jadaun2013,Liu2014,Lu2014,Shiozaki2014,Alexandradinata14,Fang2015,Shiozaki2015,Zhang2015,Chiu2016,Dong2016,Lu2016,Wieder17} and protect semimetallic phases.\cite{Young2012,Matsuura2013,Chiu2014,Shiozaki2014,Yang2014,Steinberg2014,Chiu2015,Kim2015,Yang2015,Young2015,Bradlyn2016,Chan2016,Watanabe16,Wieder2016,Wieder2016b,Zhao2016}
Most recently, groups of bands have been classified by their irreducible representations (irreps) at all high-symmetry points in the Brillouin zone.\cite{Slager2012,Kruthoff2016,Po2017}
{As noted in Ref~\onlinecite{Po2017}, irreps are a sufficient, but not necessary, condition to diagnose topological phases.
We explore the relationship between this type of classification and our classification in Appendix~\ref{sec:comparison}, using the Kane-Mele model of graphene as an example.}

Our approach goes beyond existing works by introducing a homotopic definition of equivalence; i.e., two sets of bands are topologically equivalent not only if they have the same irreps at all high-symmetry points, but also if and only if they can be smoothly deformed into each other without breaking any symmetries. The latter requirement preserves all Wilson loop invariants.\cite{ArisCohomology}
Classifications of topological crystalline materials can also be obtained using $K$-theory.\cite{KitaevClassify,Freed2013,Read2016,Shiozaki2015,Shiozaki2017}. While our method shares some phenomenological similarities to the K-theory approach, we emphasize that we have very different goals: 
instead of attempting to enumerate the topological classes mathematically permitted with a given set of symmetries, we instead derive how topological phases arise from atomic orbitals in physical systems.

Our paper is organized as follows:
in Sec~\ref{sec:definitions} we review the terminology of crystal lattices and derive how the symmetry of local orbitals determines the symmetry of the entire group of bands originating from those orbitals.
In Sec~\ref{sec:ebrs} we introduce the elementary band representation (EBR) and derive the conditions under which bands originating from a set of orbitals are elementary.
We then prove in Sec~\ref{sec:topological} that disconnected elementary band representations are topological.
Last, in Sec~\ref{sec:TR}, we introduce time reversal symmetry and derive the conditions under which bands originating from local orbitals are both elementary and time reversal symmetric.
{In a companion paper\cite{MaterialsPaper}, we develop several applications of these results to find topological materials.}


\section{From atomic orbitals to band representations}
\label{sec:definitions}

To begin, we review the concepts necessary to define a band representation, as was introduced by Zak in Ref~\onlinecite{Zak1980}. Here, we start by following the canonical reference on space groups, Ref~\onlinecite{ITA}, and then follow the derivation of a band representation from Ref~\onlinecite{Evarestov1997}. 
In order to adopt a constructive, chemistry-friendly approach to the problem, we organize the discussion to show how a local description (or, mathematically, a site-symmetry group representation) of atomic orbitals induces a global description of the band structure that determines a local $\mathbf{k} \cdot \mathbf{p}$ description at every point in momentum space.

\subsection{Wyckoff positions and stabilizer groups}
A crystal structure consists of an arrangement of atoms that is described by a Bravais lattice and which is invariant under a group of symmetry operations,
the space group (SG), $G$, of the crystal. 
We denote an element $g\in G$ that acts in real space by $\mathbf{r} \rightarrow R\mathbf{r}+\mathbf{v}$ by $\{R|\mathbf{v} \}$; {the Bravais lattice translations are denoted $\{E|\mathbf{t}\}$}.

We use $\mathbf{q}$ to denote a position in the unit cell, whether occupied by an atom or not.
A crystal with an atom at $\mathbf{q}$ must also have an atom at each site in the orbit of $\mathbf{q}$, $\{ g\mathbf{q} | g\in G\}$.
\begin{defn}\label{def:symmsite}
The set of symmetry operations, $g\in G$, that leave the site $\mathbf{q}$ fixed is called the {\bf stabilizer group} or {\bf site-symmetry group} of $\mathbf{q}$, and is denoted $G_\mathbf{q} \equiv \{ g | g\mathbf{q} = \mathbf{q} \} \subset G$.
\end{defn}
\noindent 
The site-symmetry group, $G_\mathbf{q}$, can include elements $\{ R|\mathbf{v} \}$ with $\mathbf{v}\neq \mathbf{0}$.
Nonetheless, a site-symmetry group is, by its definition, always isomorphic to a crystallographic point group.

\begin{figure}[h]
\centering
\subfloat[]{
	\includegraphics[width=1in]{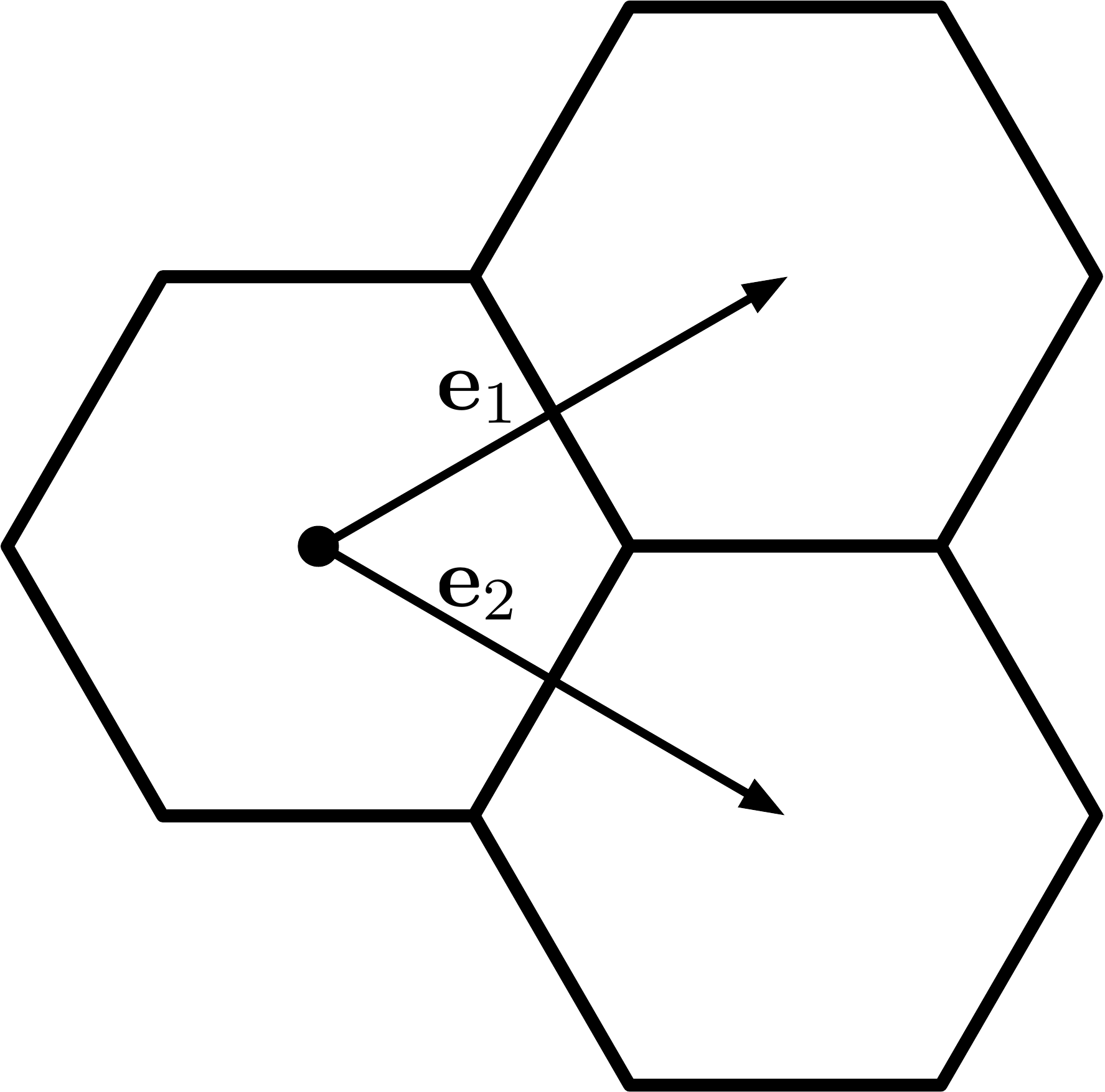}
	\label{fig:graphenebasisvectors}
}
\hspace{.1in}
\subfloat[]{
	\includegraphics[width=1.25in]{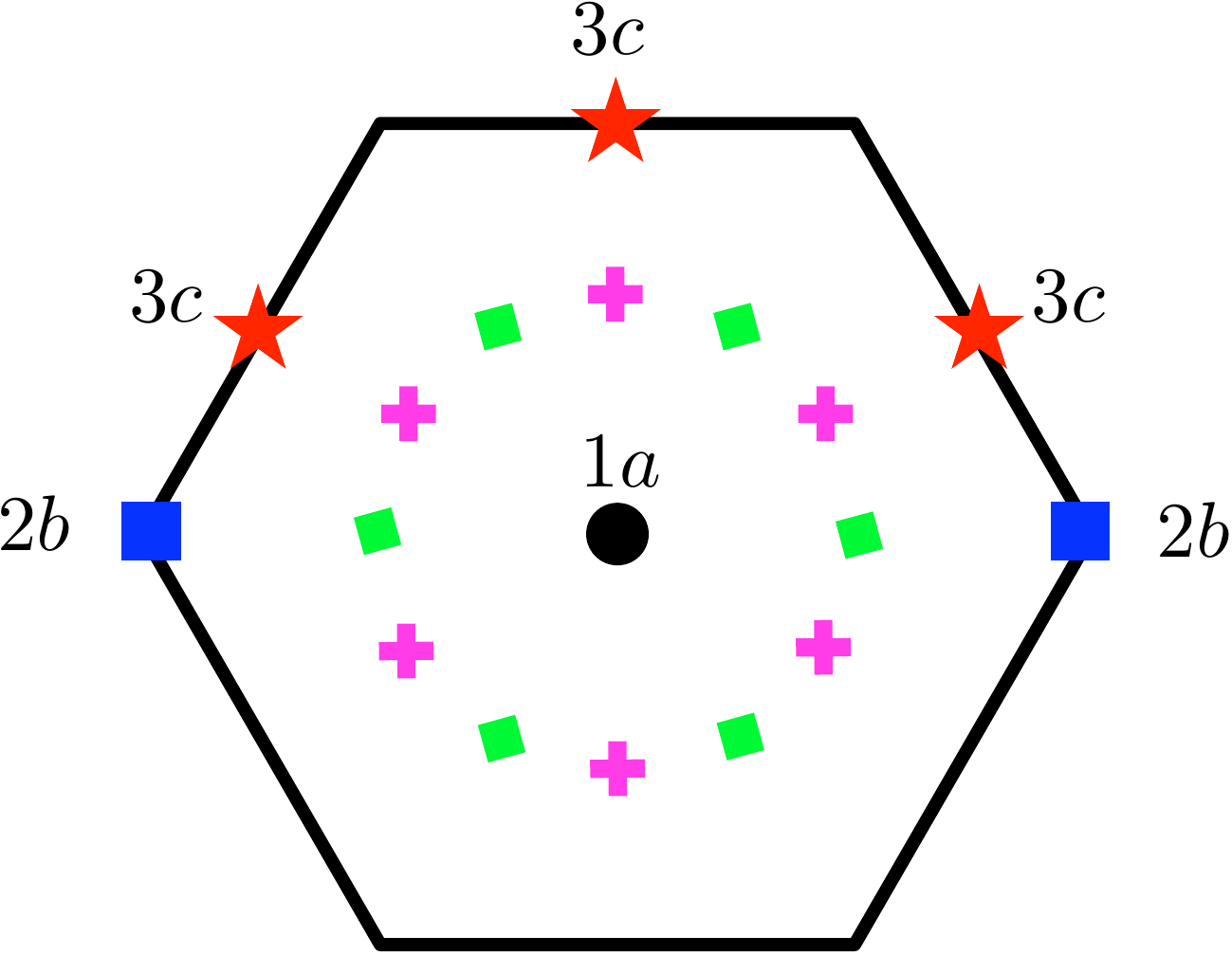}
	\label{fig:grapheneWyckoff}
}
\caption{Lattice basis vectors (a) and Wyckoff positions (b) of the hexagonal lattice. The (maximal) $1a$, $2b$ and $3c$ Wyckoff positions are indicated by a black dot, blue squares, and red stars, respectively. The {non-maximal} $6d$ and $6e$ positions are indicated by purple crosses and green squares, respectively. The multiplicity is determined by the index of the stabilizer group with respect to the point group $C_{6v}$ ($6mm$). The general position $12f$, corresponding to the orbit of a generic point, is not explicitly indicated.}
\label{fig:graphenefig}
\end{figure}

As an often-used example, we consider the two-dimensional plane group $p6mm$, which is generated by $\{ C_{3}| \mathbf{0} \}, \{ C_{2}| \mathbf{0} \}, \{ m_{1\bar{1}}| \mathbf{0} \}$ and translations, and which describes the honeycomb lattice, are shown in Fig~\ref{fig:grapheneWyckoff}.
Now consider the site $\mathbf{q} = (\mathbf{e}_1 -\mathbf{e}_2)/2$.
The mirror operation $\{ m_{11} | \mathbf{0} \}$, which is a reflection across the line perpendicular to the $\mathbf{e}_1 + \mathbf{e}_2$ axis, {(i.e., $\{m_{11} | \mathbf{0} \}$ sends $\mathbf{e}_1 + \mathbf{e}_2 \rightarrow - (\mathbf{e}_1 + \mathbf{e}_2)$)}
leaves $\mathbf{q}$ invariant, as does a $\pi$ rotation about the origin followed by a translation by $\mathbf{e}_1 - \mathbf{e}_2$.
Hence, $G_\mathbf{q}$ is generated by $\{m_{11}|\mathbf{0} \}$ and $\{C_{2}| 1\bar{1} \}$ and is isomorphic to the point group $C_{2v}$.

The site-symmetry groups of any two points in the orbit of $\mathbf{q}$ are conjugate to each other and are hence isomorphic. More generally,

\begin{defn}\label{def:Wyckoff}
{Any} two sites whose site-symmetry groups are conjugate are said to lie in the same {\bf Wyckoff position.}
Given a site in the Wyckoff position, the number of sites in its orbit that lie in a single unit cell defines the {\bf multiplicity} of the position.
\end{defn}
\noindent We always define the lattice translations relative to the primitive (not conventional) unit cell.
The Wyckoff positions of $p6mm$ are shown in Fig~\ref{fig:graphenefig}.

Given a site, $\mathbf{q}$, that is part a Wyckoff position of multiplicity $n$, we label the points in the orbit of $\mathbf{q}$ that lie in the same unit cell as $\mathbf{q}$ by $\mathbf{q}_\alpha$, where $\alpha = 1, \dots, n$ and $\mathbf{q}_1 \equiv \mathbf{q}$. For each $\alpha > 1$
there exists an element $g_\alpha \notin G_\mathbf{q}$, which is not a pure lattice translation, such that $g_\alpha \mathbf{q} = \mathbf{q}_\alpha$.
The stabilizer group of $\mathbf{q}_\alpha$ is given by 
\begin{equation}
G_{\mathbf{q}_\alpha} \equiv \{ g_\alpha hg_\alpha^{-1} | h\in G_\mathbf{q}\}
\label{eq:stabilizer}
\end{equation}
The $g_\alpha$ furnish the following coset decomposition of $G$:
\begin{equation}
G=\bigcup_{\alpha=1}^{n}g_\alpha(G_\mathbf{q}\ltimes\mathbb{Z}^3),
\label{eq:coset}
\end{equation}
where $\mathbb{Z}^3$ is the group of Bravais lattice translations and $g_1$ is the identity element. The $\ltimes$ denotes the semidirect product: $G_\mathbf{q}\ltimes\mathbb{Z}^3$ is the symmorphic space group 
{which contains the elements of $G_\mathbf{q}$
and which has the same Bravais lattice as $G$.}

We again consider $p6mm$ and use the site $\mathbf{q} = (\mathbf{e}_1 - \mathbf{e}_2)/2$ as an example.
Since $\{ C_{6}| \mathbf{0} \}\notin G_\mathbf{q} \cong C_{2v}$, the other two sites in the orbit of $\mathbf{q}$ in the unit cell are given by $\mathbf{q}_2 \equiv \{ C_{6}| \mathbf{0} \}\mathbf{q}$ and $\mathbf{q}_3 \equiv \{ C_{6}| \mathbf{0} \}^{-1} \mathbf{q}$; the red stars in Fig~\ref{fig:grapheneWyckoff} indicate the three sites.
All other symmetry operations in $p6mm$ acting on $\mathbf{q}$ take it to a position that differs from one of these sites by a pure lattice translation.

It will be important in what follows to understand how each site symmetry group, $G_\mathbf{q}$, fits into the space group, $G$. To this end, we define:
\begin{defn}\label{def:maximal}
A site-symmetry group is {\bf non-maximal} if there exists a finite group $H\neq G_\mathbf{q}$, such that $G_\mathbf{q} \subset H \subset G$. A site-symmetry group that is not non-maximal is {\bf maximal}.
A Wyckoff position containing $\mathbf{q}$ is maximal if the stabilizer group $G_\mathbf{q}$ is maximal.
\end{defn}
\noindent 
{A word of caution: if $G_\mathbf{q}\cong P$ and $G_{\mathbf{q}'}\cong P'$, where $P$ and $P'$ are abstract point groups, it is possible for $P\subset P'$ even though $G_\mathbf{q}\not\subset G_{\mathbf{q}'}$.
For example, in $P6mm$, taking $\mathbf{q} = (\mathbf{e}_1 - \mathbf{e}_2)/2$, $G_\mathbf{q}\cong C_{2v}$, while $G_\mathbf{0}\cong C_{6v}$. Even though $G_\mathbf{q}\not\subset G_\mathbf{0}$ (because, for example, $\{C_2|1\bar{1}\} \in G_\mathbf{q}$ and $\{C_2|1\bar{1}\} \not\in G_\mathbf{0}$), $C_{2v}\subset C_{6v}$.}

We can quickly find the maximal Wyckoff positions of $p6mm$ by using a sufficient, although not necessary, condition for a site-symmetry group, $G_{\mathbf{q}}$, to be maximal.
{The condition is the following:} if $\mathbf{q}$  is the unique point which is fixed by each operation in $G_\mathbf{q}$, i.e., there does not exist a second point, $\mathbf{q}' \neq \mathbf{q}$, which is also fixed by each element of $G_\mathbf{q}$, then $G_\mathbf{q}$ is maximal (we derive this condition in Appendix~\ref{sec:nopointfixed}.) 
{Since in two dimensions, rotations about the axis perpendicular to the plane keep only a single point fixed,}
any site-symmetry group which contains a rotation about an axis perpendicular to the plane is a maximal site-symmetry group.
The Wyckoff positions of $p6mm$ are shown in Fig~\ref{fig:grapheneWyckoff}. Since the $1a$, $2b$ and $3c$ positions are invariant under $6$-, $3$-, and $2$-fold rotations about the $\hat{z}$ axis, respectively, these three positions are maximal.
The non-maximal positions, $6d$ and $6e$, lie on mirror planes; they are non-maximal because their site-symmetry group is contained in those of the two maximal positions that lie on the same mirror plane ($1a$ and $3c$ or $1a$ and $2b$).

\subsection{Induction}

Suppose $n_q$ orbitals reside on the site $\mathbf{q}$, which belongs to a Wyckoff position of multiplicity $n$.
The wavefunctions of these orbitals transform under an $n_q$-dimensional representation, $\rho$, of the site-symmetry group, $G_\mathbf{q}$.
If the wavefunctions represent spinless electrons then $\rho$ will be a single-valued representation of $G_\mathbf{q}$; if the the wavefunctions are spinful, then $\rho$ will be a double-valued representation ($s=1/2$), such that a $2\pi$ rotation results in a phase of $e^{2\pi i s}=-1$.
For now, $\rho$ may be reducible or irreducible; we later show that we only need concern ourselves with irreducible representations.
On an equivalent site, $\mathbf{q}_\alpha$, the orbitals transform under the conjugate representation defined by $\rho_\alpha(h) = \rho(g_\alpha^{-1}h g_\alpha)$
for each $h\in G_{\mathbf{q}_\alpha}$; Eq~\ref{eq:stabilizer} shows that $h\in G_{\mathbf{q}_\alpha}$ implies that $g_\alpha^{-1}h g_\alpha\in G_\mathbf{q}$ for $g_\alpha \mathbf{q} = \mathbf{q}_\alpha$.

The $n_q$ orbitals on site $\mathbf{q}$ can be described by a set of Wannier functions, $W_{i1}(\mathbf{r})$, $i=1,...,n_q$, localized on $\mathbf{q}$.
For each $g\in G_\mathbf{q}$, the functions transform as
\begin{equation} gW_{i1}(\mathbf{r}) = \left[\rho(g)\right]_{ji}W_{j1}(\mathbf{r})
\label{eq:Wanniersitesym}
\end{equation}
Without loss of generality, choose the sites $\mathbf{q}_\alpha$ to be in the same unit cell as $\mathbf{q}$. Then the Wannier functions localized on $\mathbf{q}_\alpha$ are defined by $W_{i\alpha}(\mathbf{r}) = g_\alpha W_{i1}(\mathbf{r}) = W_{i1}(g_\alpha^{-1}\mathbf{r})$, where $\alpha = 1,..., n$ and $n$ is the multiplicity of the Wyckoff position.
The Wannier functions on other unit cells are defined by $\{E|\mathbf{t}_\mu \} W_{i\alpha}(\mathbf{r}) =W_{i\alpha}(\mathbf{r}-\mathbf{t}_\mu)$, where $\mathbf{t}_\mu$ is a Bravais lattice vector.
As shown in Appendix~\ref{sec:Wannierbasis}, the $n\times n_q\times N$ functions $W_{i\alpha}(\mathbf{r}-\mathbf{t}_\mu)$, where $N \rightarrow \infty$ is the number of unit cells in the system, {are closed under the symmetries of the} full space group $G$.

\subsection{Local to global}

We define the Fourier transformed Wannier functions:
\begin{equation}
a_{i\alpha}(\mathbf{k},\mathbf{r})=\sum_\mu e^{i\mathbf{k}\cdot \mathbf{t}_\mu} W_{i\alpha}(\mathbf{r}-\mathbf{t}_\mu)
\label{eq:fourier}
\end{equation}
This exchanges our infinite $n\times n_q\times N$-dimensional basis for a finite $n\times n_q$ basis for each of the $N$ $\mathbf{k}$'s in the first Brillouin zone, corresponding to $n\times n_q$ energy bands.
The induced representation in momentum space is defined as follows\cite{Zak1980}; we derive it from {the action of the space group elements} on the real space Wannier functions in Appendix~\ref{sec:Wannierbasis}:
\begin{defn}
The {\bf band representation} $\rho_G$, induced from the $n_q$-dimensional representation, $\rho$, of the site-symmetry group, $G_\mathbf{q}$, of a particular point, $\mathbf{q}$, whose orbit contains the sites $\{ \mathbf{q}_\alpha \equiv g_\alpha \mathbf{q} \}$ in the unit cell, is defined by the action
\begin{eqnarray}
&(\rho_G(h)a)_{i\alpha}(\mathbf{k},\mathbf{r})=e^{-i(R\mathbf{k})\cdot\mathbf{t}_{\beta\alpha}} \times \nonumber  \\ & { \sum_{i'=1}^{n_\mathbf{q}}}\rho_{i'i}(g_{\beta}^{-1}\{E|-\mathbf{t}_{\beta\alpha}\}hg_\alpha) a_{i'\beta}(R\mathbf{k},\mathbf{r}),
\label{eq:inducedrep}
\end{eqnarray}
for each $h=\{R|\mathbf{v}\} \in G$, where for each choice of $\alpha$ the index $\beta$ is determined by the unique coset of $G$ in Eq~(\ref{eq:coset}) that contains $hg_\alpha$:
\begin{equation}
hg_\alpha =\{E|\mathbf{t}_{\beta\alpha}\}g_{\beta}g
\label{eq:defbeta} 
\end{equation}
for some $g\in G_\mathbf{q}$, {coset representative $g_\beta$}, and Bravais lattice vector $\mathbf{t}_{\beta\alpha}$.
\end{defn}

The choice of coset representatives, $g_\alpha$, must be kept fixed throughout the construction.
The translation $\mathbf{t}_{\beta\alpha}$ is found as follows:
by moving $g_\alpha$ to the right-hand-side of Eq~(\ref{eq:defbeta}), it is evident that 
$h\mathbf{q_\alpha} = \{E|\mathbf{t}_{\beta\alpha}\}g_{\beta}g g_\alpha^{-1} \mathbf{q}_\alpha = \{E|\mathbf{t}_{\beta\alpha}\}g_{\beta}g\mathbf{q} = \{E|\mathbf{t}_{\beta\alpha}\}g_{\beta}\mathbf{q} = \{E|\mathbf{t}_{\beta\alpha}\}\mathbf{q}_\beta$ (the second and fourth equalities follow from the definition of $\mathbf{q}_{\alpha,\beta}$ and the third equality follows from $g \in G_\mathbf{q}$), which yields
\begin{equation}
\mathbf{t}_{\beta\alpha}=h\mathbf{q}_\alpha-\mathbf{q}_{\beta}. \label{eq:Rab}
\end{equation}

The matrix form of $\rho_G(h)$ consists of infinitely many $(n\cdot n_q)\times (n\cdot n_q)$ blocks. Each block is labelled by a pair $(\mathbf{k}',\mathbf{k})$, where $\mathbf{k}'$ is a row index and $\mathbf{k}$ is a column index, and corresponds to a mapping between Fourier transformed Wannier functions labelled by these $\mathbf{k},\mathbf{k}'$.
For each $h=\{R|\mathbf{v}\} \in G$ and each set of columns corresponding to $\mathbf{k}$, there is exactly one non-zero block, which corresponds to $\mathbf{k}' = R\mathbf{k}$.
We denote this block by $\rho_G^\mathbf{k}(h)$, whose matrix elements are given by,
\begin{equation}
\rho_G^\mathbf{k}(h)_{j\beta, i\alpha} \equiv e^{-i(R\mathbf{k})\cdot\mathbf{t}_{\beta\alpha}} \rho_{ji}( g_{\beta}^{-1}\{E|-\mathbf{t}_{\beta\alpha}\}hg_\alpha)
\label{eq:blocks}
\end{equation}
The full set of matrices $\rho_G^\mathbf{k}(h)$, for each $\mathbf{k}$ in the first BZ, contain all of the non-zero elements of $\rho_G(h)$ and thus completely determine the band representation.

\subsection{Global to $\mathbf{k}\cdot\mathbf{p}$}
\label{sec:globalkp}

For each $\mathbf{k}$ in the first BZ, the little group of $\mathbf{k}$, $G_\mathbf{k}$, is defined by $G_\mathbf{k} = \{ h = \{ R | \mathbf{v} \}  | R \mathbf{k} \equiv \mathbf{k}, h \in G \}$, where the equivalence relation $R\mathbf{k} \equiv \mathbf{k}$ is defined by equality up to a reciprocal lattice vector.
{$G_\mathbf{k}$ is infinite because if $h\in G_\mathbf{k}$, the operation of $h$ followed by any Bravais lattice translation is also in $G_\mathbf{k}$.}
The set $\{ \rho_G^\mathbf{k}(h) | h \in G_\mathbf{k} \}$ furnishes an $(n\cdot n_q)\times (n\cdot n_q)$ representation of $G_\mathbf{k}$, whose matrix elements are given by Eq~(\ref{eq:blocks}).
We denote this representation by $\rho_G \downarrow G_\mathbf{k}$; {this is a subduction of $\rho_G$ onto $G_\mathbf{k}$, projected onto the Wannier functions at $\mathbf{k}$}.
Although the little group {$G_\mathbf{k}$} is infinite, the representation of two space group operations, $\{ R |\mathbf{v}\}$ and $\{R | \mathbf{v}+ \mathbf{t}_1\}$, which differ by a full lattice translation, $\mathbf{t}_1$, will differ only by an overall phase $e^{-i (R\mathbf{k})\cdot \mathbf{t}_1}=e^{-i \mathbf{k}\cdot \mathbf{t}_1}$ in $\rho_G \downarrow G_\mathbf{k}$. Hence, $\rho_G\downarrow G_\mathbf{k}$ is a ``small representation.''\cite{Cracknell}


The characters of $\rho_G \downarrow G_\mathbf{k}$ are given by, {for $h\in G_\mathbf{k}$},
\begin{equation}
\chi_G^\mathbf{k}(h) \equiv \sum_\alpha e^{-i(R\mathbf{k})\cdot \mathbf{t}_{\alpha\alpha}}\tilde{\chi} \left[\rho(g_\alpha^{-1}\{E | -\mathbf{t}_{\alpha\alpha} \}hg_\alpha ) \right],
\label{eq:character}
\end{equation} 
where 
\begin{equation}
\tilde{\chi}[\rho(g)] = \begin{cases} \chi[\rho(g)] & \text{if }g \in G_\mathbf{q} \\ 0 & \text{if }g\notin G_\mathbf{q}
\end{cases}
\end{equation}
and $\chi[\rho(g)]$ denotes the character of the matrix representative of $g$ in the representation $\rho$.

We would like to know how many times each irrep, $\sigma_i^\mathbf{k}$, of $G_\mathbf{k}$ appears in $\rho_G \downarrow G_\mathbf{k} $, i.e., we would like to find the coefficients, $m_i^\mathbf{k}$, which satisfy 
\begin{equation}
(\rho\uparrow G)\downarrow G_\mathbf{k}\cong \bigoplus_im_i^\mathbf{k}\sigma_i^\mathbf{k},
\end{equation} 
where we have used $\cong$ to denote the equivalence of representations and introduced the shorthand
\begin{equation}
m_i\sigma_i\equiv\underbrace{\sigma_i\oplus\sigma_i\oplus\dots\oplus\sigma_i}_{m_i}.
\end{equation}
The multiplicities, $m_i^\mathbf{k}$, are determined by the linear independence and completeness of the characters: they are the unique solution to the set of equations,
\begin{equation}
\chi^\mathbf{k}_G(h)=\sum_i m_i^\mathbf{k} \chi_{\sigma_i}^\mathbf{k}(h),\, \forall \,\, h\in G_\mathbf{k},
\label{eq:decompose}
\end{equation}
where $\chi_{\sigma_i}^\mathbf{k}(h)$ denotes the characters of $\sigma_i^\mathbf{k}$.

This general formalism explains how energy bands in momentum space inherit their properties from the real-space orbitals on Wyckoff positions in the unit cell.
An example of how to compute the characters $\chi_G^\mathbf{k}(h)$ is given in Appendix~\ref{sec:charactersgraphene}.

\subsection{Example: induction from $1a$}
\label{sec:inductionexample}

We now show how to induce a band representation from the site $\mathbf{q} = (0,0)$ in the $1a$ Wyckoff position in $p6mm$.
All operations in $p6mm$ of the form $\{ R| \mathbf{0} \}$ leave this position invariant; thus, the site-symmetry group, $G_\mathbf{q}$, is generated by $\{ C_{3}| \mathbf{0} \}, \{ C_{2}| \mathbf{0} \} $ and $\{ m_{1\bar{1}} | \mathbf{0} \}$; $G_\mathbf{q}$ is isomorphic to $C_{6v}$.
The character table for the irreps of $C_{6v}$ is shown in Table~\ref{table:c6v}.\cite{GroupTheoryPaper}

\begin{table}[h]
\begin{tabular}{c|ccccccc}
Rep & $E$ & $C_{3z}$ & $C_{2z}$ & $C_{6z}$ & $m$ & $C_{6z}m$ & $\bar{E}$ \\
\hline
$\Gamma_1$ &1 &1 &1 &1 &1 &1 &1 \\
$\Gamma_2$ &1 &1 &1 &1 &-1 &-1 &1 \\
$\Gamma_3$ &1 &1 &-1 &-1 &-1 &1 &1 \\
$\Gamma_4$ &1 &1 &-1 &-1 &1 &-1 &1 \\
$\Gamma_5$ &2 &-1 &2 &-1 &0 &0 &2 \\
$\Gamma_6$ &2 &-1 &-2 &1 &0 &0 &2 \\
${\bar\Gamma}_7$ &2 &-2 &0 &0 &0 &0 &-2 \\
${\bar\Gamma}_8$ &2 &1 &0 &-$\sqrt{3}$ &0 &0 &-2 \\
${\bar\Gamma}_9$ &2 &1 &0 &$\sqrt{3}$ &0 &0 &-2 \\
\end{tabular}
\caption{The character table for the group $C_{6v}$. (In this table and hereafter each of the conjugacy classes is represented by a symmetry operation belonging to the class; conjugacy classes whose members are obtained
from those listed combined with $\bar{E}$ are not shown, e.g. $\bar{C}_{3z}=C_{3z}\bar{E}$ and $\bar{C}_{6z}=C_{6z}\bar{E}$ {are not shown}.) The irreps $\Gamma_1$--$\Gamma_6$ are all single valued, while $\bar{\Gamma}_7,\bar{\Gamma}_8,$ and $\bar{\Gamma}_9$ are double valued. $\bar{\Gamma}_7$ is the $|S=3/2,m_z=\pm 3/2\rangle$ representation, $\bar{\Gamma}_8$ is the $|S=5/2,m_z=\pm 5/2\rangle$ representation and $\bar{\Gamma}_9$ is the spin-$\half$ representation.}\label{table:c6v}
\end{table}

For each irrep, $\Gamma_{j}$ (or $\bar{\Gamma}_j$) in Table~\ref{table:c6v}, we can induce a band representation according to Eq~(\ref{eq:inducedrep}).
Since the Wyckoff position has multiplicity one, the index $\alpha$ in Eq~(\ref{eq:inducedrep}) is trivial, and we omit it.
Consequently, in Eq~(\ref{eq:defbeta}), {since there is only one coset}, $g_\alpha = g_\beta = E$ ($E$ is the identity operator) and for each $h=\{R|\mathbf{t} \}\in G$, Eq~(\ref{eq:defbeta}) simplifies to $h=\{E|\mathbf{t} \} \{ R| \mathbf{0} \}$, using the fact that $\{ R| \mathbf{0} \} \in G_\mathbf{q}$ and $\mathbf{t}_{\alpha\beta}=\mathbf{t}$. Then Eq~(\ref{eq:inducedrep}) yields the band representation:
\begin{equation}  
\left( \rho_{G,j} (\{R|\mathbf{t} \} )a\right)_i (\mathbf{k},\mathbf{r}) = e^{-i(R\mathbf{k})\cdot \mathbf{t} } \sum_{i'} \left[ \Gamma_j(R)\right]_{i'i}a_{i'}(R\mathbf{k},\mathbf{r} ),
\label{eq:induce1a}
\end{equation}
where the indices $i,i' = 1, ..., n_q = |\Gamma_j |$ and the representation dimension $|\Gamma_j |$ is exactly equal to the character of $E$ in Table~\ref{table:c6v}.
Eq~(\ref{eq:induce1a}) shows that each element, $\{R|\mathbf{t}\}$, in the space group
is represented in the band representation by an infinite matrix, due to the fact that $\mathbf{k}$ takes $N \rightarrow \infty$ values, where $N$ is the number of unit cells.
That infinite matrix transforms the Fourier transformed Wannier function at $\mathbf{k}$ to one at $R\mathbf{k}$, transforms the orbital $i$ to $i'$ with the coefficient $\left[ \Gamma_j (R)\right]_{i'i}$, and gives an overall phase $e^{-i(R\mathbf{k})\cdot\mathbf{t}}$.
It is evident that the infinite dimensional representation can be reduced into finite dimensional space group representations that act on the finite set of Fourier transformed Wannier functions $\{ a_i(R \mathbf{k}, \mathbf{r})| i = 1, ..., |\Gamma_j |, \{R|\mathbf{t}\} \in G \}$ for fixed $\mathbf{k}$\cite{Cracknell,GroupTheoryPaper}.

This procedure generalizes to Wyckoff positions with multiplicity greater than one by including the index $\alpha$ in Eq~(\ref{eq:inducedrep}). 
The only additional difficulty is that $g_{\alpha}$ and $g_\beta$ in Eq~(\ref{eq:defbeta}) are non-trivial: $\alpha$ is determined by the left-hand-side of Eq~(\ref{eq:inducedrep}) ($\mathbf{q}_\alpha = g_\alpha \mathbf{q} $) and $\beta$ must be found from the coset decomposition in Eq~(\ref{eq:coset}); $\mathbf{t}_{\alpha\beta}$ is then obtained from Eq~(\ref{eq:Rab}).
An example is shown in Appendix~\ref{sec:charactersgraphene}.


\section{Elementary band representations}\label{sec:ebrs}

We would like to determine when a band representation can be decomposed into smaller, unique, band representations.
To this end, it is necessary to define an equivalence relation of band representations, which we first introduced in Ref~\onlinecite{NaturePaper}:
\begin{defn}\label{def:equiv}
Two band representations $\rho_G$ and $\sigma_G$ are {\bf equivalent} iff there exists a unitary matrix-valued function $S(\mathbf{k},t,g)$ smooth in $\mathbf{k}$ and continuous in $t$ such that for all $g\in G$
\begin{enumerate}
\item $S(\mathbf{k},t,g)$ defines a band representation according to Eq~(\ref{eq:blocks}) for all $t\in[0,1]$,
\item $S(\mathbf{k},0,g)=\rho_G^\mathbf{k}(g)$, and
\item $S(\mathbf{k},1,g)=\sigma_G^\mathbf{k}(g)$ 
\end{enumerate}
\end{defn}
This definition implies that $\rho_G^\mathbf{k}$ and $\sigma_G^\mathbf{k}$ restrict to the same little group representations at all points in the BZ.
 However, it is necessary to have a stronger definition of equivalence because it is possible for two EBRs to have the same representations at all points in the BZ but be physically distinguishable by a Berry phase.\cite{Bacry1988b,Michel1992,Bacry1993,RegRep,NaturePaper} 
 We work out examples in Appendix~\ref{sec:ExNotEquiv} and Appendix~\ref{sec:comparison}.
 In these cases, even though $S(\mathbf{k},0,g)$ and $S(\mathbf{k},1,g)$ restrict to the same little group representations at all points in the Brillouin zone, any mapping, $S(\mathbf{k},t,g)$, between them will \emph{not} be a band representation -- it will either break a crystal symmetry or break time reversal locally -- and in this case $S(\mathbf{k},0,g)$ and $S(\mathbf{k},1,g)$ will be inequivalent.

Def~\ref{def:equiv} preserves any quantized Wilson loop\cite{ArisCohomology} invariant, which is understood as follows: since $S$ is continuous in $t$, any property of a band representation evolves continuously under the equivalence $S$. In particular, the Wilson loop matrices computed from the bands in the representation $\rho_G^{\mathbf{k}}$ {evolve continuously into} the Wilson loop matrices computed in the representation $\sigma_G^{\mathbf{k}}$. As such, two equivalent band representations cannot be distinguished by any quantized Wilson loop invariant. This includes the recent case of invariants formed from Wilson loops of Wilson loops.\cite{hughesbernevig2016}

 We now explain how to construct the homotopy utilized in Def~\ref{def:equiv}, given two distinct sites, $\mathbf{q}$ and $\mathbf{q}'$, with respective site-symmetry groups, $G_{\mathbf{q}}$ and $G_{\mathbf{q}'}$, that have intersection 
 $G_0=G_{\mathbf{q}}\cap G_{\mathbf{q}'}$, where $G_0$ is itself a stabilizer group of some site, $\mathbf{q}_0$.\cite{Bacry1988}
Then $\mathbf{q}_0$ has a free parameter that interpolates between $\mathbf{q}$ and $\mathbf{q}'$;
for example, if $\mathbf{q}, \mathbf{q}'$ are high-symmetry points, then $\mathbf{q}_0$ describes a line that connects them.
Given any representation, $\sigma$, of $G_0$, we can induce a band representation in two different ways, either $(\sigma\uparrow G_{\mathbf{q}})\uparrow G$ or $(\sigma\uparrow G_{\mathbf{q}'})\uparrow G$, which are equivalent.
Then, the free parameter in $\mathbf{q}_0$ is exactly the parameter $t$ in Def~\ref{def:equiv} that continuously tunes the band representation between $\mathbf{q}$ and $\mathbf{q}'$. 
This establishes a sufficient condition for equivalence:
\begin{prop}
Given two sites, $\mathbf{q} \neq \mathbf{q}'$, and representations $\rho$ and $\rho'$ of $G_\mathbf{q}$ and $G_{\mathbf{q}'}$, respectively,
the band representations $\rho \uparrow G$ and $\rho' \uparrow G$ are equivalent if there exists a site $\mathbf{q}_0$ and representation $\sigma$ of $G_{\mathbf{q}_0}$ such that {$G_{\mathbf{q}_0}\subset \left(G_\mathbf{q} \cap G_{\mathbf{q}'}\right)$}, $\rho = \sigma \uparrow G_\mathbf{q}$ and $\rho' = \sigma \uparrow G_{\mathbf{q}'}$.
\label{prop:equiv}
\end{prop} 
\noindent
Sufficient and necessary conditions for equivalence are established by combining Prop~\ref{prop:equiv} with the fact that equivalence is transitive (which follows from Def~\ref{def:equiv}.)

Using Definition \ref{def:equiv} and following Ref~\onlinecite{NaturePaper}, we define:
\begin{defn}
A band representation is called {\bf composite} if it is equivalent to the direct sum of other band representations. A band representation that is not composite is called {\bf elementary}.
\label{def:composite}
\end{defn}

We will now identify all the elementary band representations associated to a given space group. We first derive two necessary but not sufficient conditions for a band representation to be elementary when time-reversal symmetry is ignored. We then add time-reversal symmetry, and discuss its effects on the theory of band representations. By construction, all band representations admit a description in terms of localized Wannier functions, as they are induced from the representation of some site-symmetry group $G_\mathbf{q}$ (associated to the space group $G$) under which the wavefunctions of local orbitals transform.

First, because induction commutes with direct sums, i.e.
\begin{equation}
(\rho_1\oplus\rho_2)\uparrow G = (\rho_1\uparrow G)\oplus (\rho_2\uparrow G)
\label{eq:reducible}
\end{equation}
we deduce that 
\begin{conditions}
reducible representations of $G_\mathbf{q}$ induce composite band representations.
\label{cond:reducible}
\end{conditions}  
\noindent From this we conclude that we need only examine the irreps of the stabilizer groups in order to enumerate all EBRs.
Second, induction is transitive: given groups $K\subset H\subset G$, and a representation $\rho$ of $K$, it follows that
\begin{equation}
(\rho\uparrow H)\uparrow G = \rho\uparrow G.
\label{eq:inducesubgroup}
\end{equation}
From this it follows that 
\begin{conditions}
all elementary band representations can be induced from irreducible representations of the maximal site symmetry groups.
\label{cond:EBRirreps}
\end{conditions} 
\noindent This reduces the search for EBRs to bands induced from the maximal Wyckoff positions (c.f.  Def~\ref{def:maximal}).

\subsection{Exceptions}\label{sec:exceptions1}

However, there are cases where an irrep of the site-symmetry group of a maximal Wyckoff position induces a composite band representation.
This can happen because the decomposition of an infinite dimensional representation into elementary representations is not necessarily unique.\cite{Bacry1988}
Given a maximal Wyckoff position $\{ \mathbf{q} \}$, and an irrep, $\rho$, of $G_\mathbf{q}$, $\rho_G$ will be equivalent to a composite band representation induced from a different maximal Wyckoff position $\{\mathbf{q}'\}$ if there exists
\begin{enumerate}
\item a path, $l$, which connects $\mathbf{q}$ and $\mathbf{q}'$, such that the site-symmetry group of each point in $l$ is equal to $G_0 \equiv G_\mathbf{q}\cap G_{\mathbf{q}'}$, and
\item a representation, $\sigma$, of $G_0$ such that $\rho=\sigma\uparrow G_\mathbf{q}$ is irreducible, while $\sigma\uparrow G_{\mathbf{q}'}$ is reducible.
\end{enumerate}
{Then $\rho_G$ is equivalent to the composite band representation $(\sigma\uparrow G_{\mathbf{q}'})\uparrow G$.}
The equivalence is furnished by inducing the band representations from each point in $l$, using the irrep, $\sigma$. Since every point in the line $l$ has the same site-symmetry group, this indeed gives an equivalence as per Def.~\ref{def:equiv}. We give an example of such an equivalence in Sec~\ref{sec:exceptionexample}.  
We will refer to the irreps of maximal site symmetry groups that do not induce EBRs as \emph{exceptions}.
Band representations induced from irreps of maximal site symmetry groups that are equivalent to exceptions are also exceptions.

We now describe how to determine which irreps of maximal site symmetry groups induce exceptions. This is a crucial step towards our goal of enumerating all EBRs, because such irreps do \emph{not} induce EBRs, even though they satisfy the necessary (but not sufficient) Condition~\ref{cond:EBRirreps} below Eq~(\ref{eq:inducesubgroup}).
To do this, we first note that all single-valued point group representations have dimension $1,2$ or $3$, while all double-valued representations have dimension $1,2$ or $4$ (only the single and double cubic point groups have irreps of dimension $3$ and $4$, corresponding to the (single-valued) vector spin-$1$ and (double-valued) spin-$3/2$ irrep respectively). Next, given a representation, $\sigma$, of some $G_0=G_{\mathbf{q}}\cap G_{\mathbf{q}'}$, the induced representation $\rho=\sigma\uparrow G_{\mathbf{q}}$ will have dimension
\begin{equation}
\mathrm{dim}(\rho)=\mathrm{dim}(\sigma)[G_{\mathbf{q}}:G_0]\in \mathrm{dim}(\sigma)\mathbb{Z},
\label{eq:induceddim}
\end{equation}
where $[G_{\mathbf{q}}:G_0]$ denotes the index of $G_0$ as a subgroup of $G_{\mathbf{q}}$, which, by Lagrange's theorem, is always a positive integer greater than one, if $G_0 \neq G_\mathbf{q}$. In order for $\rho$ both to be an irrep of $\Gq$ and to be equivalent to an induction of a representation $\sigma\uparrow G_{\mathbf{q}}$, it must have dimension larger than $1$. Hence,  $\mathrm{dim}(\rho)=2$ or $3$ for single valued representations, or $\mathrm{dim}(\rho)=2,4$ for double valued representations. 
We now focus on the double-valued group representations; the single-valued exceptions were considered in Refs~\onlinecite{Bacry1988,Michel2001}. 
{Since $\mathrm{dim}(\sigma)$ must divide $\mathrm{dim}(\rho)$, we deduce that either $\mathrm{dim}(\sigma)=1, \mathrm{dim}(\rho)=2$ or $4$ or $\mathrm{dim}(\sigma)=2, \mathrm{dim}(\rho)=4$.}
We show in Appendix~\ref{sec:irrepsdim4} that in the groups where $G_{\mathbf{q}}$ has a four-dimensional irrep and an index two subgroup $G_0$, that there is no site $\mathbf{q}_0$ which has $G_0$ as its site-symmetry group.
{Thus, exceptions can only occur when $\mathrm{dim}(\sigma)=1$.}
{After enumerating all point group triplets $\Gq,\Gqp, G_0$, where $\Gq, \Gqp$ are maximal subgroups of $G$; $G_0=\Gq\cap\Gqp$; and $[\Gq : G_0]=2$ or $4$,}
we have found all $1D$ irreps of $G_0$ which induce irreps of $\Gq$, and reducible reps of $\Gqp$ and matched these cases to site-symmetry groups of maximal Wyckoff positions for the $230$ space groups.\cite{NaturePaper}

For the classical space groups (those that permit only single-valued representations), the list of all Wyckoff positions for which these exceptions occur are tabulated in Ref.~\onlinecite{Bacry1988,Michel2001}, which we repeat for convenience in Table \ref{table:sbr}; we have computed the analogous list for the double space groups, shown in Table \ref{table:dbr}, which we first presented in Ref~\onlinecite{NaturePaper}. We have thus established:
\begin{prop}
A band representation, $\rho_G$, is elementary if and only if it can be induced from an irreducible representation $\rho$ of a maximal site-symmetry group $G_\mathbf{q}$, and if it is not {listed in Tables~\ref{table:sbr} or \ref{table:dbr}.}
\label{prop:EBR}
\end{prop} 
\noindent
{Thus, an algorithmic listing of all band representations\cite{GraphTheoryPaper,GraphDataPaper} does not need to include the irreps in Tables~\ref{table:sbr} or \ref{table:dbr}.}
Band representations induced from the site-symmetry groups $G_\mathbf{q}$ listed in these tables are composite. They reduce into a sum of elementary band representations induced from $G_{\mathbf{q}'}$, listed in the second column of Table~\ref{table:sbr} or \ref{table:dbr}.
 We give an example of such a band representation in Appendix~\ref{sec:exceptionexample}.
Prop~\ref{prop:EBR} and Tables~\ref{table:sbr} and \ref{table:dbr} accomplish our goal of identifying \emph{all} the EBRs associate with a given SG.

\section{Band Connectivity and Topological Systems}
\label{sec:topological}

We have so far established the conditions under which a band representation induced from an irrep of the site-symmetry group of a maximal Wyckoff position is elementary.
In this section, we establish the connection between EBRs and topological bands, which we define as follows:

\begin{defn}
A set of bands are in the {\bf atomic limit} of a space group if they can be induced from localized Wannier functions consistent with the crystalline symmetry of that space group. Otherwise, they are {\bf topological}.
\label{def:atomic}
\end{defn}
Band representations, by their construction, describe a system in the atomic limit.\cite{Evarestov1997}
Thus, topological bands must be groups of bands that satisfy the crystal symmetry in momentum space, but nevertheless do not transform as a band representation.
{In other words, they cannot be induced from localized Wannier orbitals that obey the crystal symmetry.\cite{Soluyanov2011}}

\subsection{Compatibility relations and quasi band representations}
\label{sec:compatibility}

A set of Bloch wavefunctions that obey the crystal symmetry will, at each point in the BZ, transform as a sum of irreps of the little group at that point.  However, the irreps at each point in the BZ cannot be chosen independently.\cite{GraphTheoryPaper,GraphDataPaper,NaturePaper}
In particular, given a high-symmetry line emanating from a high-symmetry point, the little group of the line is a subgroup of the little group of a point. It follows that each irrep that appears in the band decomposition at the point can be {subduced} to a sum of irreps that appear on the line; in this way, the irreps along the line are completely determined by the irreps that appear at the point. This decomposition is referred to as a ``compatibility relation'' between the high-symmetry point and line.\cite{Cracknell} Compatibility relations also exist for planes and volumes emanating from lines and planes, respectively.

Every band representation yields a solution to the compatibility relations by construction. On the other hand, there exist solutions to the compatibility relations that are not band representations. Following Bacry\cite{Bacry1993} and Ref~\onlinecite{NaturePaper}, where we also explored some of these ideas, we define
\begin{defn}
A {\bf quasi band representation (qBR)} is any solution to the compatibility relations.
\end{defn}

\noindent As we mentioned above, band representations describe the atomic limit.
The reverse is also true: any set of atomic orbitals induces a band representation.
We are thus motivated to define:
\begin{defn}
\label{def:tqbr}
A qBR that is not a (composite or elementary) band representation is a {\bf topological quasi band representation (tqBR)}.
\end{defn}
Because they are not band representations, tqBRs cannot describe bands with localized, crystal-symmetric Wannier functions: if they existed, such Wannier functions would reside on some Wyckoff position and transform under a representation of the site-symmetry group of that position, thereby inducing a band representation.
This is the natural extension to crystal-symmetric systems of the results by Soluyanov and Vanderbilt\cite{Soluyanov2011,Soluyanov2012}, which showed that $\mathbb{Z}_2$ topological insulators lack time-reversal symmetric Wannier functions.

\subsection{Connectivity of EBRs}
\label{sec:connectivity}

Now let us consider a Hamiltonian, $\mathcal{H}$, constructed from localized orbitals whose eigenstates transform in an elementary band representation $\rho_G$ associated to a space group $G$; $G$ may be a classical, double, or even magnetic space group in any number of dimensions. 
Next, assume that the energy bands corresponding to the representation $\rho_G$ can be divided into two disconnected components, which are separated by an energy gap, $\Delta$, {which can vary as a function of $\mathbf{k}$, but which is always finite.}
Let $P_1(P_2)$ be the projector onto the disconnected group of bands with lower(higher) energy.
Then $P_{1}(P_2)$ commutes with all the symmetry generators $g$ in $G$.
Thus, the projected Hamiltonian $\mathcal{H}_1 \equiv P_1 \mathcal{H} P_1$ commutes with all the symmetries of $G$.
Now suppose that the non-zero eigenstates of $P_1$ and $P_2$ transform according to band representations, $\rho_G^{(1)}$ and $\rho_G^{(2)}$, respectively, induced from a set of orbitals that transform into each other under the symmetries of $G$.
This implies  $\rho_G=\rho_G^{(1)} \oplus \rho_G^{(2)}$, which contradicts the hypothesis that $\rho_G$ is elementary.
Thus, there are two possibilities: either the bands that correspond to $\rho_G$ are connected or the non-zero eigenstates of $P_1$ or $P_2$ do not both transform like a band representation of $G$ and hence cannot be derived from a set of orbitals that transform into each other under the crystal symmetries, i.e., they do not correspond to a symmetry-preserving atomic limit.
We consider the latter case to be topological.
We conclude,
\begin{prop}\label{prop:connectivity}
All elementary band representations are either connected (as an energy graph), or (if disconnected) yield at least one group of bands that is a (weak, strong, or crystalline) topological insulator.
\end{prop} 
Ref.~\onlinecite{PoFragile} provides an example of a Hamiltonian where an EBR splits into two groups of bands separated by an energy gap such that one of the two groups of bands allows for symmetric, localized Wannier functions, while the irreps that appear in the other group of bands forbid their existence.
This possibility was overlooked in Ref.~\onlinecite{NaturePaper}, although it is contained in our theory of elementary band representations.
It follows from Proposition~\ref{prop:connectivity} that
\begin{cor}\label{cor:connectivity}
Any isolated set of bands that is not equivalent to a band representation (composite or elementary) gives a strong, weak, or crystalline topological insulator.
\end{cor}

We conclude from Corollary \ref{cor:connectivity} that when tqBRs occur in the spectrum of a Hamiltonian, that Hamiltonian is in a topological phase. 
This is a band property, independent of where the Fermi level sits in a particular system.
In addition, Corollary \ref{cor:connectivity} is much more powerful than the existing ad hoc approach to computing topological crystalline invariants: even without knowledge of a particular invariant, it determines whether a set of isolated bands is topological.
{Furthermore, a list of distinct tqBRs would themselves define a topological index.}

It also follows from Proposition~\ref{prop:connectivity} that tqBRs arise following a topological metal-to-insulator phase transition, where {single} connected elementary band representation becomes disconnected {into two or more tqBRs.} Similarly, {tqBRs }can occur in a phase transition between a topologically trivial and topologically nontrivial insulator when a gap closes in a composite band representation made up of two (without loss of generality) elementary band representations and re-opens it to  give rise to two tqBRs. This possibility, where two elementary band representation combine and give rise to two tqBRs, is not followed further: although our theory {identifies and} characterizes those situations as topological, additional {quantitative} information is necessary to predict them -- those topological situations occur if the spin-orbit coupling is stronger than some critical value and are hence quantitative by nature. In contrast, the disconnected EBRs where two tqBRs form one EBR, is much stronger: the tqBRs are topological irrespective of the {quantitative parameters of the model}.  

The preceding logic leads us to one of the most important consequences of this work: we can identify candidate topological crystalline insulator (TCI) phases by forming all possible solutions to the compatibility relations\cite{GroupTheoryPaper} and then looking for disconnected energy graphs that  are not EBRs. We develop this search further in forthcoming work.\cite{GraphDataPaper,GraphTheoryPaper,NaturePaper}

\subsection{Obstructed atomic limit}
\label{sec:obstructed}

We remarked (below Def~\ref{def:tqbr}) that tqBRs do not have crystal-symmetric Wannier functions.
Yet, ``topological insulators'' in one dimension present no obstruction to the formation of symmetric localized Wannier functions\cite{Kohn59,Kivelson1982}. Similarly, the subclass of weak topological phases in two and three dimensions that inherit their topology from one-dimensional systems also allow for a Wannier-description\cite{Read2016}.  These cases are considered ``topological'' because they display a quantized polarization invariant.  Similarly, the quadrupole insulators proposed in Ref~\onlinecite{hughesbernevig2016} in higher dimensions, even though not decomposable into one-dimensional wires, also have crystal symmetric Wannier states, despite being different from the {trivial} atomic limit (in that they exhibit a quantized quadrupole moment).
In the nontrivial state, symmetric, localized Wannier functions exist, but do not reside on the atomic sites and, further, cannot be continued back to the atomic sites without either closing the gap to other bands or breaking {a} symmetry.

Since these phases possess symmetric, localized Wannier functions, they can be continuously deformed to \emph{an} atomic limit; however, this limit does not describe the position of the ions.
Hence, {these phases} describe hybridization transitions.
This is very different from the tqBRs we defined in Def~\ref{def:tqbr}, which cannot be continuously deformed to any atomic limit.
In Ref~\onlinecite{NaturePaper} we proposed the following definition to distinguish these two cases:
\begin{defn}
A set of bands is in the {\bf obstructed atomic limit} when they possess symmetric, localized Wannier functions that reside on a Wyckoff position distinct from the Wyckoff position of the underlying ions and which cannot be smoothly deformed to the ionic position.
\end{defn}  A specific example of this situation was discussed in Sec V of Ref~\onlinecite{NaturePaper}.

\subsection{How to determine whether a set of bands is a band representation}
\label{sec:determinebandrep}

From Corollary~\ref{cor:connectivity}, we know that an isolated set of bands is topological if it is not equivalent to a band representation. We now seek to answer {the following question} on a practical level: given an isolated set of bands, how does one determine whether they are equivalent to a band representation?

As explained below Def~\ref{def:equiv}, the notion of equivalence preserves the set of irreps that appear at each high-symmetry point in the BZ and any quantized Wilson loop invariant. 
While the latter are difficult to compute -- a full list of all Wilson loop invariants is not enumerated anywhere in the literature -- the former is straight-forward.

Thus, a practical route to determining whether a set of bands, $\mathcal{B}$, is \emph{not} a band representation is as follows:
{first, enumerate all EBRs for the particular space group and list the irreps that appear in each EBR at each high-symmetry point.\cite{NaturePaper,GroupTheoryPaper}}
Next, compute the irreps at each high-symmetry point for the bands in $\mathcal{B}$.
If the set of irreps that have been computed for the bands in $\mathcal{B}$ cannot be obtained from a linear combination of the EBRs in the space group, then the bands in $\mathcal{B}$ do not comprise a band representation and, by Corollary~\ref{cor:connectivity}, are topological.

If the irreps that appear in $\mathcal{B}$ can be obtained from a linear combination of the EBRs of the space group, 
{then one must compute symmetric and localized Wannier functions for the bands in $\mathcal{B}$ to confirm that they are equivalent to the atomic limit defined by the linear combination of EBRs or compute a Berry phase that will distinguish the two.}
This is because, as shown in Appendices~\ref{sec:comparison} and \ref{sec:ExNotEquiv} {(motivated by examples in Refs~\onlinecite{Bacry1988b,Michel1992,Bacry1993})}, it is possible for two distinct groups of bands to have the exact same irreps at all high-symmetry points, but different Berry phases (recall, this is exactly why we require the homotopic notion of equivalence, as in Def~\ref{def:equiv}.)
{If the orbitals and atoms that contribute to $\mathcal{B}$ are known, this information can be sufficient to
exclude the existence of a homotopy between the band representation induced from the orbitals that contribute to $\mathcal{B}$ and the linear combination of EBRs if there does not exist a symmetry-preserving path along which their corresponding atomic orbitals can be continuously deformed into each other}; i.e., an equivalence between the band representations is forbidden according to Prop.~\ref{prop:equiv}.
An example is discussed at the end of Appendix~\ref{sec:ExNotEquiv}.

We note, as shown in Appendix~\ref{sec:comparison}, that it is possible that the irreps that appear at high-symmetry points in the valence bands can be obtained from a linear combination of EBRs while those in the conduction bands cannot; in this case, the conduction bands must be topological by Corollary~\ref{cor:connectivity}.

\section{Time reversal symmetry}
\label{sec:TR}

In a time reversal invariant system, the Wannier functions must respect time-reversal symmetry in real space. For spinless systems this means the Wannier functions must either be real or come in complex-conjugate pairs. For spinful systems, the Wannier functions must always come in spin up and spin down pairs. 
{We now characterize band representations in the presence of time reversal symmetry.}
We will see that imposing time reversal symmetry affects the properties of band representations in both real space and momentum space.

\subsection{Physically irreducible representations}
\label{sec:physicalirreps}

Mathematically, the Wannier functions at a site $\mathbf{q}$ will obey local time-reversal symmetry precisely when they transform according to a time-reversal invariant representation of the site-symmetry group, $G_\mathbf{q}$. 
Let $\rho$ denote an irrep of the site-symmetry group.
To determine whether $\rho$ is time-reversal invariant requires computing the Frobenius-Schur indicator (reviewed in Appendix~\ref{sec:Frobenius}), which labels $\rho$ as real, quaternionic or complex.
If $\rho$ is real and single-valued or quaternionic and double-valued, then there exists an anti-unitary time reversal operator that squares to $+1$ or $-1$, respectively.
In any other case, $\rho$ is not time-reversal symmetric.
{Then, }to restore time reversal symmetry in real space, $\rho$ must be paired with its complex conjugate, $\rho^*$.

Representations that cannot be decomposed as a sum of other time reversal-preserving representations are commonly referred to as \emph{physically irreducible}.\cite{Bilbao1}
Thus, if $\rho$ is an irrep of $G_\mathbf{q}$ which is real and single-valued or quaternionic and double-valued, then it is physically irreducible.
Otherwise, $\rho\oplus \rho^*$ is physically irreducible: even though $\rho\oplus \rho^*$ is a reducible representation of $G_\mathbf{q}$ {(without TR)}, it cannot be decomposed into irreps that respect time reversal symmetry.

We will later want to know which point group irreps are time reversal invariant; to this end, we have computed the Frobenius-Schur indicator (Eq.~(\ref{eq:frobenius})) for all representations of all $32$ point groups (as tabulated in Ref.~\onlinecite{Cracknell}) and found the following:
\begin{enumerate}
\item {All point group irreps with dimension greater than one are either real and single-valued or quaternionic and double-valued, except for six complex irreps (two of the three double-valued irreps of $T$ and four of the six double-valued irreps of $T_h$).}
\item The one-dimensional double-valued irreps are either real or complex (consequently, they are never time reversal invariant, which constitutes Kramers theorem.)
\item The one-dimensional single-valued irreps are either real or complex.
\end{enumerate}

\subsection{Time reversal symmetric band representations}

Band representations induced from a time-reversal invariant representation of the site-symmetry group will be endowed with a time-reversal symmetry operator, which can be found by generalizing the induction procedure in Eq~(\ref{eq:inducedrep}), as follows: let $\rho(T)$ denote the anti-unitary representative of the time reversal operator; $\rho(T)$ is the product of a unitary matrix and the complex conjugation operator, $K$.
Since time reversal commutes with all space group operations, {it is does not mix Wannier functions on different sites, i.e.}, in Eq~(\ref{eq:defbeta}), $\alpha = \beta$ and $\mathbf{t}_{\beta\alpha}=0$; consequently, Eq~(\ref{eq:inducedrep}) yields the band representation of the time reversal operator:
\begin{equation} 
(\rho_G(T) a)_{i\alpha}(\mathbf{k},\mathbf{r})= \sum_{i'=1}^{n_\mathbf{q}} \rho_{i'i}(T)a_{i'\alpha}(-\mathbf{k},\mathbf{r})
\label{eq:inducedrepTR}
\end{equation}

We will refine our definition (Def.~\ref{def:composite}) of an EBR in the presence of time reversal symmetry.
Following the logic of Sec.~\ref{sec:ebrs}, we first define
\begin{defn}\label{def:pequiv}
Two band representations $\rho_G^\mathbf{k}$ and $\sigma^\mathbf{k}_G$ are {\bf physically equivalent} if they are equivalent (in the sense of Def.~\ref{def:equiv}), and if, for all $t$, the homotopy $S(\mathbf{k},t,g)$ between them (c.f. Def.~\ref{def:equiv}) is a band representation induced from a sum of {some} time-reversal invariant site-symmetry group representations.
\end{defn}
In other words, physically equivalent band representations are related by a homotopy that preserves real-space time reversal symmetry. Generalizing Def.~\ref{def:composite} for elementary band representations, we then define\cite{NaturePaper} 

\begin{defn}\label{def:pelem}
A band representation is {\bf physically elementary} iff it is induced from a (locally) time-reversal invariant representation of a site-symmetry group, and if it is not physically equivalent to a direct sum of other band representations. Otherwise, a band representation induced from a locally time-reversal invariant representation of a site-symmetry group is {\bf physically composite}.
\end{defn}

In other words, physically elementary band representations (pEBRs) are the building blocks for band structures which respect time-reversal symmetry in momentum space, and whose Wannier functions respect time-reversal symmetry locally in real space.

\subsection{Exceptions}
\label{sec:exceptions2}

According to Def~\ref{def:pelem}, physically elementary band representations are induced from time reversal invariant representations of maximal site-symmetry groups.
Because Condition~\ref{cond:reducible} also applies in the presence of time reversal symmetry, we further deduce that pEBRs are induced from physically irreducible representations of maximal site-symmetry groups.
However, a physically irreducible representation of a maximal site symmetry group does not always induce a physically elementary band representation (the induced band representation will always be time-reversal symmetric, per Eq~(\ref{eq:inducedrepTR}), but it might be composite.)
This phenomenon was discussed without time reversal symmetry in Sec~\ref{sec:exceptions1} {and resulted in Tables~\ref{table:sbr} and \ref{table:dbr}.}
We now consider the conditions under which physically irreducible site-symmetry group representations induce pEBRs.

If $\rho$ is a time reversal symmetric irrep of a site-symmetry group and $\rho \uparrow G$ is an EBR, then $\rho \uparrow G$ will also be a pEBR {by definition} (for, suppose not: then $\rho \uparrow G$ would be physically equivalent to a direct sum of other band representations, which violates the assumption that it is an EBR).
Thus, real and single-valued or quaternionic and double-valued irreps (i.e., irreps that are time reversal symmetric) that induce EBRs will also induce pEBRs.

Consequently, there are two situations in which a physically irreducible representation of a maximal site-symmetry group, $G_\mathbf{q}$, may induce a composite physical band representation: 
first, if the physically irreducible representation of $G_\mathbf{q}$, $\rho$, is also an irrep of $G_\mathbf{q}$ and $\rho \uparrow G$ is a composite band representation, {i.e., an exception.} This is exactly the mechanism described in Sec~\ref{sec:exceptions1}  and the cases where this can occur are listed in Tables~\ref{table:sbr} and \ref{table:dbr}; we describe how these lists change with time reversal symmetry in Sec~\ref{sec:exceptionsTR1}.
The second is when the physically irreducible representation is not an irrep of $G_\mathbf{q}$, in which case the physically irreducible representation is of the form $\rho\oplus \rho^*$, where $\rho$ is an irrep of $G_\mathbf{q}$. This is a generalization of the mechanism in Sec~\ref{sec:exceptions1}, which we detail in Sec~\ref{sec:exceptionsTR2}.

\subsubsection{When an irrep of a maximal site-symmetry group is time reversal invariant, but induces a physically composite band representation}
\label{sec:exceptionsTR1}

Here we consider the case where an irrep, $\rho$, of a maximal site-symmetry group, $G_\mathbf{q}$, is time reversal symmetric.
We proved in the previous section that $\rho \uparrow G$ can only fail to be a pEBR if it fails to be an EBR, which can only happen for the irreps listed in Table~\ref{table:sbr} or \ref{table:dbr}.
The exceptional band representations appearing in Tables~\ref{table:sbr} and \ref{table:dbr} are precisely those band representations induced from irreps of $G_\mathbf{q}$ that are equivalent to composite {band representations induced from the site-symmetry group $G_{\mathbf{q}'}$}. In all cases, the equivalence $S(\mathbf{k},t)$ is via band representations induced from \emph{one-dimensional} representations of the lower-symmetry group $G_{\mathbf{q}_0}$ (we proved this in Sec.~\ref{sec:exceptions1}.)
Thus, the homotopy between an exceptional band representation at position $\{\mathbf{q}\}$ and a composite band representation at position $\{\mathbf{q}'\}$ has Wannier functions localized on a line with site-symmetry group $G_0$ and transforming in a one-dimensional representation. 
To determine whether that one-dimensional representation respects time reversal symmetry, we distinguish between the single-valued (spinless) and double-valued (spinful) group representations.
As explained in Sec~\ref{sec:physicalirreps}, one-dimensional double-valued site-symmetry representations necessarily break time reversal symmetry in real space; thus, this homotopy violates time reversal symmetry in real space for the double-valued groups.
Consequently, {none of} the spinful exceptions listed in Table~\ref{table:dbr} are \emph{physically equivalent} to composite representations.
We conclude that if $\rho$ is a double-valued irrep of $G_\mathbf{q}$ and $\rho$ is time reversal symmetric, then $\rho \uparrow G$ is {always} a pEBR.

Unlike the spinful case, a one-dimensional spinless representation can be time reversal invariant if it is real.
We have checked that this is the case for those exceptions with $G_0=C_{2v}$, which appear below the double-line in Table~\ref{table:sbr}. 
For all other exceptions in Table~\ref{table:sbr}, the relevant representation $\rho_0$ of $G_{\mathbf{q}_0}$ is complex and hence not time reversal invariant {(specifically, in $C_3$, $\Gamma_{2,3} \uparrow D_3 = \Gamma_3$; in $C_6$, $\Gamma_{5,6} \uparrow D_6=\Gamma_5$, while $\Gamma_{2,3} \uparrow D_6 = \Gamma_6$; and in $C_4$, $\Gamma_{3,4} \uparrow D_4 = \Gamma_5$.\cite{Bilbao3})}
Thus, the homotopy between $\rho \uparrow G$ and the composite band representation induced from a representation of $G_{\mathbf{q}'}$ breaks time reversal and is not a physical equivalence {when $G_0\neq C_{2v}$}. 
We conclude that if $\rho$ is a single-valued irrep of $G_\mathbf{q}$ and $\rho$ is time reversal symmetric, then $\rho\uparrow G$ is a pEBR unless it appears below the double-line in Table~\ref{table:sbr}.

\subsubsection{When an irrep of a maximal site-symmetry group is not time reversal invariant}
\label{sec:exceptionsTR2}

\begin{table}[b]
\begin{tabular}{cccc}
Reducing group ($G_{\mathbf{q}'}$) & SGs \\
\hline
$C_{2h}$ & $84,87,135,136$ \\
$D_2$ & $112,116,120,121,126,130,$\\
& $133,138,142,218,230$ \\
$D_4$ & $222$ \\
$D_{2d}$ & $217$ \\
$T$ & $219,228$  \\
\end{tabular}
\caption{Additional exceptional band representations with time-reversal. In all cases, the exceptional representation is the physically irreducible two-dimensional representation of {$G_\mathbf{q} = S_4$}. For the space groups listed in this table, this band representation decomposes through $G_{\mathbf{q}_0}=C_2$ into a composite band representation induced from the reducing group $G_\mathbf{q}'$. The first column gives the reducing group, while the second column gives the associated space groups for which the exception occurs.}\label{table:TRexceptions}
\end{table}

We now consider a new class of exceptions with spinless systems with time-reversal symmetry that do not appear in Table~\ref{table:sbr} (at the end of this section, we address why they do not occur in the spinful case): it may be the case that there exist sites $\mathbf{q},\mathbf{q}'$ and $\mathbf{q}_0$ with $G_\mathbf{q}\cap G_{\mathbf{q}'}=G_{\mathbf{q}_0}$, such that a real irrep, $\rho_0$ of $G_{\mathbf{q}_0}$, induces a representation $\rho\oplus\rho^*$ of $G_\mathbf{q}$, which is physically irreducible {(but reducible without TR)}, and that the induced representation $\rho_0\uparrow G_{\mathbf{q}'}$ is physically reducible.
Then the induced band representation $(\rho\oplus\rho^*)\uparrow G$ is an exception in the presence of TR because it is physically equivalent to the composite band representation, $(\rho_0\uparrow G_{\mathbf{q}'})\uparrow G$.
This situation would not be an exception without TR because, without TR, $\rho\oplus\rho^*$ is a reducible representation of $G_\mathbf{q}$ and, hence, induces a composite band representation per Eq~(\ref{eq:reducible}).

We have listed the exceptions where $\rho_0\uparrow G_{\mathbf{q}}=\rho\oplus\rho^*$ is physically irreducible but $\rho_0 \uparrow G_{\mathbf{q}'}$ is physically reducible in
Table~\ref{table:TRexceptions}. 
We now explain how to find the entries in this table: 
as noted above Eq~(\ref{eq:induceddim}), all single-valued (spinless) point group representations, $\rho$, are $1, 2$ or $3$ dimensional.
However, we explained in Sec~\ref{sec:physicalirreps} that all $2D$ and $3D$ representations are real, and hence time reversal invariant.
Thus, if $\rho\oplus\rho^*$ is a physically irreducible representation of $G_\mathbf{q}$, then ${\rm dim}(\rho) = 1$ and, consequently, ${\rm dim}(\rho \oplus \rho^*)=2$.
Since $G_0$ is a proper subgroup of $G_\mathbf{q}$, ${\rm dim}(\rho \oplus\rho^* = \rho_0 \uparrow G_\mathbf{q})> {\rm dim}(\rho_0)$; consequently, ${\rm dim}(\rho_0)=1$ {(this dimension counting explains why we do not need to consider yet another type of exception where $\rho\oplus \rho^*$ is induced from a physically irreducible representation of the form $\rho_0 \oplus \rho_0^*$, where $\rho_0$ is an irrep of $G_0$: because ${\rm dim}(\rho_0 \oplus \rho_0^*)\geq 2$ and $G_0$ is a proper subgroup of $G_\mathbf{q}$, $\rho_0 \oplus \rho_0^*$ {could} not induce a representation of $G_\mathbf{q}$ of dimension $2$.)}
For $\rho \oplus\rho^*$ to be physically irreducible, $\rho$ must be a complex irrep of $G_\mathbf{q}$ (recall from Sec~\ref{sec:physicalirreps} that there are no quaternionic $1D$ irreps). The only point groups with single-valued complex $1D$ irreps are $C_4, C_{4h}, S_4, C_3,C_{3i},C_6, C_{6i}, C_{6h}, T$ and $T_h$; we now consider these cases:
\begin{description}
\item[{  $G_\mathbf{q} \cong C_3,C_{3i},C_6, C_{6i}, C_{6h}, T$ or $T_h$}]
In these cases, $G_\mathbf{q}$ contains a three-fold rotation, $C_{3}$. 
Since $G_{\mathbf{q}_0}$ is an index-two subgroup of $G_\mathbf{q}$, $G_{\mathbf{q}_0}$ must also contain $C_{3}$.
{Since are interested in a real representation, $\rho_0$, of $G_{\mathbf{q}_0}$, $\chi^{\rho_0}(C_{3})=1$, where $\chi^\sigma(g)$ denotes the character of $g$ in the representation $\sigma$.
Since $G_{\mathbf{q}_0}$ is an index-two subgroup of $G_\mathbf{q}$, there must exist an element $h\in G, h\notin G_{\mathbf{q}_0}$ such that $h^2\in G_{\mathbf{q}_0}$.
We now deduce the character of $C_3$ in the induced representation, $\chi^{\rho_0\uparrow G}(C_3)$, using the Frobenius formula, which says that if $hC_{3}h^{-1} \notin G_{\mathbf{q}_0}$ then $\chi^{\rho_0\uparrow G}(C_3)=\chi^{\rho_0}(C_3)$, while if $hC_{3}h^{-1} \in G_{\mathbf{q}_0}$, then $\chi^{\rho_0\uparrow G}(C_3)=\chi^{\rho_0}(C_3) + \chi^{\rho_0}(hC_{3}h^{-1})$.
In the first case, $\chi^{\rho_0\uparrow G}(C_{3})=1$.
In the second case, $(hC_{3}h^{-1})^3=E$ implies $\left[\chi^{\rho_0}(hC_3h^{-1})\right]^3=1$ and since $\rho_0$ is a real representation, $\chi^{\rho_0}(hC_3h^{-1})=1$.
Consequently, in the second case, $\chi^{\rho_0\uparrow G}(C_{3})=2$.}
However, we deduced above that $\rho$ is a complex $1D$ representation; this means that $\chi^{\rho}(C_{3})=e^{\pm 2\pi i/3}$ and, consequently, $\chi^{\rho\oplus\rho^*}(C_{3})=-1$.
Thus, if $G_\mathbf{q}$ contains a three-fold rotation, the representation induced from $\rho_0$ will not be {of the form $\rho\oplus\rho^*$ where $\rho$ is a complex $1D$ irrep of $G$;} hence, it does not contribute to an exception in Table~\ref{table:TRexceptions}.

\item[$G_\mathbf{q} \cong C_4$ or $C_{4h}$]
$C_{2(h)}$ is an index two subgroup of $C_{4(h)}$.
However, we have checked on the Bilbao Crystallographic Server\cite{Bilbao2} that there is no Wyckoff position, $\mathbf{q}'$, distinct from $\mathbf{q}$ such that $G_\mathbf{q} \cap G_{\mathbf{q}'} \cong C_{2(h)}$. Thus, if $G_\mathbf{q}\cong C_{4(h)}$, it does not contribute to an exception in Table~\ref{table:TRexceptions}.

\item[$G_\mathbf{q} \cong S_4$]
We see that all entries in Table~\ref{table:TRexceptions} come from the case $G_\mathbf{q}\cong S_4$.
The only index-two subgroup of $S_4$ is $C_2$. One can easily check that the one-dimension real representation $\rho_-$ of $C_2$ with $\chi^{\rho_-}(C_{2z})=-1$ induces a two-dimensional physically irreducible representation $\rho\oplus\rho^*$ of $S_4$, with $\chi^{\rho}(IC_{4z})=i$.
To complete Table~\ref{table:TRexceptions}, one must find all space groups with distinct sites, $\mathbf{q}$, $\mathbf{q}'$ and $\mathbf{q}_0$ such that $G_\mathbf{q}\cong S_4$ and $G_\mathbf{q}\cap G_{\mathbf{q}'} = G_{\mathbf{q}_0} \cong C_2$.
If, for the representation $\rho_-$ of $G_{\mathbf{q}_0}$, $\rho_-\uparrow G_{\mathbf{q}'}$ is physically reducible 
then the space group is listed in Table~\ref{table:TRexceptions}.
\end{description}

We now address why this type of exception cannot occur for spinful systems with double-valued representations: as described in Sec~\ref{sec:physicalirreps}, the only double-valued representations $\rho \oplus \rho^*$ that are physically irreducible  representations {(but reducible when TR is not present)} of a site-symmetry group $G_\mathbf{q}$ occur when either $G_\mathbf{q}= T $ or $T_h$ or when ${\rm dim}(\rho) = 1$.
In the former case, we {considered every index-two subgroup, $G_{\mathbf{q}_0}$, of $T$ and $T_h$ and checked that an irrep of $G_{\mathbf{q}_0}$ never induces a representation $\rho \oplus \rho^*$ of $T$ or $T_h$ where $\rho$ is complex.}
In the latter case, if ${\rm dim}(\rho) = 1$, then ${\rm dim}(\rho \oplus \rho^*) = 2$.
{Thus, if there existed an irrep, $\rho_0$, of $G_{\mathbf{q}_0}$, such that $G_{\mathbf{q}_0}$ is an index-two subgroup of $G_\mathbf{q}$ and $\rho_0\uparrow G=\rho\oplus\rho^*$, then $\dim(\rho_0)=1$.}
However, as discussed in Sec~\ref{sec:physicalirreps}, one-dimensional, spinful, irreps cannot be time reversal invariant; {hence, no such $\rho_0$ exists.}

To summarize, we have the following general result:
\begin{prop}
A spinless (single-valued) band representation, $\rho_G$, is physically elementary if and only if it can be induced from a physically irreducible representation, $\rho$, of a maximal site-symmetry group $G_\mathbf{q}$, unless
either
(1) $\rho$ appears {below} the double line in Table~\ref{table:sbr} or
(2) $\rho$ is the two dimensional physically irreducible representation of $G_\mathbf{q}\cong S_4$ in a SG listed in Table~\ref{table:TRexceptions}.

A spinful (i.e. double-valued) band representation $\rho_G$ is physically elementary if and only if it can be induced from a physically irreducible representation $\rho$ of a {maximal} site-symmetry group $G_\mathbf{q}$.
\label{prop:physelementary}
\end{prop}


\subsection{Connectivity of band structures: physical topological quasi band representations}

In order to discuss time-reversal invariant topological phases, we define, in analogy with Def.~\ref{def:pelem},
\begin{defn}
A {\bf physical quasi band representation (pqBR)} is any solution to the compatibility relations, which also respects time-reversal symmetry in momentum space.
\end{defn}
and
\begin{defn}
A pqBR that is not equivalent to any sum of physically elementary band representations is a {\bf physical topological quasi band representation (ptqBR)}.
\end{defn}

In analogy to the discussion below Def~\ref{def:tqbr}, ptqBRs cannot have both time reversal and crystal-symmetric Wannier functions: if they existed, such Wannier functions would reside on some Wyckoff position and transform under a representation of the site-symmetry group of that position, thereby inducing a band representation.

It is straight-forward to generalize Proposition~\ref{prop:connectivity} and Corollary~\ref{cor:connectivity}:
\begin{prop}
All physically elementary band representations are either connected or, if disconnected, yield (weak, strong, or crystalline) topological bands
\end{prop}
and
\begin{cor}
Any isolated set of bands that is not physically equivalent to a physical band representation is a strong, weak, or crystalline topological insulator.
\label{cor:notphysicalbands}
\end{cor}

It follows that when ptqBRs occur in the spectrum of a Hamiltonian, that Hamiltonian is in a topological phase. 

We now briefly comment on one route to find ptqBRs by utilizing the exceptional band representations in Table~\ref{table:dbr}. Those band representations in Table~\ref{table:dbr} without a sharp ($\sharp$) can be realized in momentum space with disconnected components, while still respecting the compatibility relations (see Sec~\ref{sec:compatibility}) and time reversal symmetry in momentum space (by respecting time reversal symmetry in momentum space, we mean that for each irrep of the little group $G_\mathbf{k}$ that appears at $\mathbf{k}$, its complex conjugate representation appears at $G_\mathbf{-k}$). 
Note, importantly, that this does \emph{not} imply that time-reversal symmetry is respected in real space and has a matrix representation in the sense of Eq.~(\ref{eq:inducedrepTR}).
However, we also know from the discussion following Def~\ref{def:pequiv} that these ptqBRs are \emph{not} physical band representations: each disconnected component is distinguishable from any physically elementary band representation. Hence, the exceptional band representations in Table~\ref{table:dbr} are ptqBRs.
Thus, Table~\ref{table:dbr} serves as a list of space groups (and particular Wyckoff positions) to search for candidate TCI materials.

Physical topological quasi band representations can also be found if $\rho \oplus \rho^*$ is physically irreducible (and not listed as an exception in Table~\ref{table:TRexceptions}), but $\rho \uparrow G$ (and thus also $\rho^*\uparrow G$) is time reversal invariant in momentum space.
In this case, there will generically be an energy gap between bands induced from the band representation $\rho\uparrow G$ and those induced from $\rho^*\uparrow G$. Thus, $\rho\uparrow G$ and $\rho^*\uparrow G$ describe two sets of connected bands that do not admit crystal and time-reversal symmetric Wannier functions, since the band representations $\rho\uparrow G$ and $\rho^*\uparrow G$ do not respect time-reversal symmetry in real space and so are not separately physical band representations.
It follows from Corollary~\ref{cor:notphysicalbands} that the gap between these two band groups is (crystalline) topological.\cite{Soluyanov2011,Read2016}
Note, however, that it is not always the case that $\rho \uparrow G$ is time reversal invariant in momentum space: the other possibility is that $\rho\uparrow G$ and $\rho^*\uparrow G$ transform into one another under time reversal symmetry; in this case, the two band reps are forced to be degenerate at the TRIM points in the BZ, and together form a connected physically elementary band representation.

\section{Accidental degeneracies}

The band structure of a particular Hamiltonian might include bands transforming under different elementary band representations that overlap in energy.
Taking inspiration from Herring\cite{Herring37}, we refer to these bands as {\bf accidentally connected} because symmetry does not require them to be connected. It follows from the preceding arguments that we can remove the accidental connections by adding to the Hamiltonian a (potentially large) perturbation which respects all crystal symmetries. However, one should not use this as an excuse to dismiss the importance of accidental connections: the non-uniqueness of the decomposition of composite band representations means that accidental connections may be physically interesting. For instance, when the stabilizer group $G_\mathbf{q}$ of a non-maximal Wyckoff position is a subgroup of \emph{two} different maximal stabilizers $G_\mathbf{q'}$ and $G_\mathbf{q''}$, the composite band representations induced from $G_\mathbf{q}$ can be reduced in two equivalent ways: either into EBRs induced from $G_\mathbf{q'}$ or from $G_\mathbf{q''}$. The connectivities of the band representations in these reductions can be different, and perturbing the Hamiltonian can drive a transition between different band connectivities. The transition region (if it represents a phase rather than a critical point) will be dominated by an accidental connection of these band representations. In real space, this process can be visualized as moving the centers of the Wannier orbitals of the crystal from Wyckoff position $\{\mathbf{q}'\}$ to Wyckoff position $\{\mathbf{q}''\}$, along a line with stabilizer group $G_\mathbf{q}$.  
We have worked an example for a one-dimensional chain of $s$ and $p$ orbitals with inversion symmetry -- i.e., the Su-Schrieffer-Heeger\cite{ssh1979} or Rice-Mele\cite{RiceMele} model -- in Ref~\onlinecite{NaturePaper}.

\section{Conclusions}

In this work, we provided the theoretical framework of our re-introduction\cite{Zak1980} of EBRs as a natural way to determine the topological properties of bands. The main idea, presented in Ref~\onlinecite{NaturePaper}, is that, because {EBRs} unify the real and momentum space descriptions of a crystalline solid, they can describe both trivial and topological behavior. In particular, we showed that disconnected EBRs yield bands that lack a local real space description that preserves crystal (and/or time reversal) symmetry and hence are topological.

The connection to real space will also be useful to find topological materials: namely, by searching for materials whose orbitals at the Fermi level induce disconnected EBRs.
Similarly, semi-metal can be found by searching for materials whose connected EBRs at the Fermi level will be partially filled.

In addition, we have shown that all of the EBRs in a particular space group can be generated by induction from irreps of maximal site-symmetry groups. 
This significantly reduces the amount of work necessary to enumerate all EBRs in the space group, a task that we take on in the related Refs.~\onlinecite{GraphDataPaper} and \onlinecite{GraphTheoryPaper}.
This result makes possible a systematic search for topological materials.\cite{MaterialsPaper}

\begin{acknowledgements}
BB and JC thank M. Zaletel and Judith H{\"o}ller for fruitful discussions. 
MGV would like to thank Gonzalo Lopez-Garmendia for help with computational work. 
{BB, JC and BAB thank Adrian Po and Ashvin Vishwanath for helpful discussions.}
BB, JC, ZW, and BAB acknowledge the hospitality of the Donostia International Physics Center, where parts of this work were carried out. 
JC acknowledges the hospitality of the Kavli Institute for Theoretical Physics, and BAB acknowledges the hospitality and support of the \'{E}cole Normale Sup\'{e}rieure and Laboratoire de Physique Th\'{e}orique et Hautes Energies. The work of MGV was supported by FIS2016-75862-P and FIS2013-48286-C2-1-P national projects of the Spanish MINECO. The work of LE and MIA was supported by the Government of the Basque Country (project IT779-13)  and the Spanish Ministry of Economy and Competitiveness and FEDER funds (project MAT2015-66441-P). 
{ZW and BAB acknowledge the support of the NSF EAGER Award DMR -- 1643312, ONR - N00014-14-1-0330, ARO MURI W911NF-12-1-0461 and NSF-MRSEC DMR-1420541, which were used to develop the initial theory and for futher ab initio work. The development of the practical part of the theory, tables, and some of the code-development was funded by Department of Energy de-sc0016239, the Simons Investigator Award, the Packard Foundation and the Schmidt Fund for Innovative Research.}
\end{acknowledgements}

\section{Tables}
\label{sec:tables}

\onecolumngrid

\begin{table}[h]
\begin{tabular}{cccccc}
{Irrep}  & Site symm. grp. & Reducing grp. &  Intersection grp. & Rep dim. & SGs \\
($\rho$) & ($G_\mathbf{q}$) & ($G_{\mathbf{q}'}$) & ($G_0$) &  &  \\
\hline
$\Gamma_3(E)$ & $D_3$ & $C_{3i}$& $C_3$& $2$ & $163, 165, 167, 228, 230$ \\
& & $T_h$&$C_3$ &$2$ & $223$ \\
& & $O$ &$C_3$ &$2$ &$211$ \\
&  & $T$ &$C_3$ &$2$ & $208, 210, 228$ \\
&  & $C_{3h}$ &$C_3$ & $2$ & $188, 190, 192, 193$\\ 
 \hline
$\Gamma_{5,6}(E_{2,1})$ & $D_6$ &$C_{6h}$ &$C_{6}$ &$2$ & $192$ \\
\hline
 $\Gamma_5(E)$ & $D_4$ &$O$ & $C_{4}$& $2$ & $207,211,222$ \\
&  &$C_{4h}$ & $C_{4}$&$2$ & $124,140$ \\
  \hline
  \hline
$\Gamma_5(E)$ & $D_{2d}$ &$D_{4h}$ &$C_{2v}$ &$2$ & $229$ \\
&  &$T_h$ &$C_{2v}$ &$2$ & $226$ \\
& & $T_d$ &$C_{2v}$ &$2$ & $215,217,224$ \\
& & $D_{2h}$ &$C_{2v}$ &$2$ &$131, 132, 139, 140, 223$
\end{tabular}
\caption{{Single-valued irreps of maximal site-symmetry groups that yield \emph{composite} band representations} and thus do not need to be considered in a search for EBRs; computed by Bacry, Michel, and Zak.\cite{Bacry1988,Michel2001} Point group symbols are given in Schoenflies notation.\cite{Cracknell} 
{Irreps are listed in the notation of Ref~\onlinecite{Cracknell} and, parenthetically, the notation of Ref~\onlinecite{Mulliken33}.
The first column gives the irrep of the maximal site-symmetry group, $G_\mathbf{q}$, listed in the second column. 
This irrep induces a composite band representation.}
The third column gives the site-symmetry group, $G_{\mathbf{q}'}$, into whose band representations this composite representation can be reduced. The fourth column gives the intersection group, $G_0=G_{\mathbf{q}}\cap G_{\mathbf{q}'}$. The fifth column gives the dimension of the irrep of $G_\mathbf{q}$ which induces the composite band rep. The sixth column indicates the space groups for which this occurs by their sequential numbers.
{Groups that appear below the double line are also physically equivalent to a composite band rep, while those above the double line are not; see Proposition~\ref{prop:physelementary} and surrounding text.}
}\label{table:sbr}
\end{table}
\begin{table}[h]
{\renewcommand{\arraystretch}{1.15}
\begin{tabular}{cccccc}
{Irrep} & Site symm. grp. & Reducing grp. &  Intersection grp. & Rep dim. & SGs \\
($\rho$) & ($G_\mathbf{q}$) &   ($G_{\mathbf{q}'}$) & ($G_0$) &  &  \\
\hline
$\bar{\Gamma}_8(\bar{F})$ & $T_d$ & $D_{3d}$ &$C_{3v}$ & $4$ & $224, 227$ \\
&  & $O_h$ & $C_{3v}$ & $4$ & $225$\\
\hline
$\bar{\Gamma}_6(\bar{E}_1)$ & $D_3$ & $T_h$& $C_3$ & $2$ & $223$\\
& &$O$ &$C_3$ & $2$ & $211$ \\
& &$T$ &$C_3$ & $2$ &$208,210,228$ \\
& & $C_{3h}$ &$C_3$ &$2$ &$188^\sharp, 190^\sharp, 192, 193$\\
& & $C_{3i}$ &$C_3$ &$2$& $163^\sharp, 165^\sharp, 167^\sharp, 228^\sharp, 230^\sharp$\\
 \hline
$\bar{\Gamma}_7(\bar{E}_3)$ &$D_{3h}$ & $D_{3d}$ &$C_{3v}$ & $2$ & $193^\sharp,194^\sharp$ \\
\hline
$\bar{\Gamma}_9(\bar{E}_1)$ & $D_6$ &$C_{6h}$ &$C_6$ &$2$ & $192^\sharp$\\
\hline
$\bar{\Gamma}_{6,7}(\bar{E}_{2,1})$ & $D_4$ & $O$ &$C_4$ &$2$ & $207,211, 222$\\
& & $C_{4h}$ &$C_4$ &$2$ &$124^\sharp,140^\sharp$\\
 \hline
$\bar{\Gamma}_5(\bar{E})$&$C_{2v}$ &$C_{6v}$ &$C_s$  &$2$ & $183$\\
& &$C_{3v}$ &$C_s$ & $2$ & $183$\\
& &$C_{2h}$ &$C_s$ & $2$ & $51^\sharp,  63^\sharp,  67^\sharp,  74^\sharp, 138^\sharp$\\
& & $C_{4v}$ &$C_s$ &$2$&$99, 107$ \\
& & $D_{2d}$ & $C_s$ & $2$ & $115,137$ \\
 \hline
$\bar{\Gamma}_6(\bar{E})$& $D_{2}$ &$T$& $C_2$&$2$ & $195,197, 201, 208, 209, 218$\\
& & $D_6$ &$C_2$ &$2$&$177,192$\\
& & $D_3$ &$C_2$ & $2$ & $177, 192, 208, 211, 214, 230$\\
& & $S_{4}$ &$C_2$ &$2$&$112^\sharp, 116^\sharp, 120^\sharp, 121, 126, 130^\sharp, 133^\sharp, 138^\sharp, 142^\sharp, 218, 230$ \\
& & $D_{2d}$ &$C_2$ &$2$& $111, 121, 132, 134, 224$ \\
& &$C_{2h}$ &$C_2$ &$2$ & $49^\sharp,  66^\sharp,  67^\sharp,  69,  72^\sharp, 124, 128, 132, 134, 135^\sharp, 138^\sharp, 192$ \\
& & $D_4$ &$C_2$ & $2$ & $89,  97, 124, 126, 211$ \\
& & $D_{3d}$ & $C_2$ & $2$ & $224$ \\
& & $O$ & $C_2$ & $2$ & $209$
\end{tabular}
}
\caption{{Double-valued irreps of maximal site-symmetry groups that yield \emph{composite} band representations.
Irreps are listed in the notation of Ref~\onlinecite{Cracknell} and, parenthetically, the notation of Ref~\onlinecite{Mulliken33}.
The first column gives the irrep, $\rho$, of the maximal site-symmetry group,  $G_\mathbf{q}$, listed in the second column.
This irrep induces a composite band representation.} Point groups symbols are given using Schoenflies notation\cite{Cracknell}; for example $C_s$ is the point group generated by a single mirror. The third column gives the site-symmetry group, $G_{\mathbf{q}'}$, into whose band representations this composite representation can be reduced. The fourth column gives the intersection group, $G_0=G_{\mathbf{q}}\cap G_{\mathbf{q}'}$. The fifth column gives the dimension of the irrep of $G_\mathbf{q}$ which induces the composite band rep. The sixth column indicates the space groups for which this occurs by their sequential number.
A sharp ($\sharp$) indicates that while the band representation is disconnected in momentum space when time reversal symmetry is ignored, 
{it is forced to be connected when time reversal symmetry is included (note the refinement with respect to the asterisks in Ref.~\onlinecite{NaturePaper})}\cite{GraphDataPaper,progbandrep}.}
\label{table:dbr}
\end{table}

\twocolumngrid

\clearpage

\appendix

\section{Proof that a site-symmetry group with exactly one fixed point is maximal}
\label{sec:nopointfixed}

In this appendix we prove that a sufficient condition for a site-symmetry group, $G_\mathbf{q}$, to be maximal, as defined in Def~\ref{def:maximal}, is that $\mathbf{q}$ is the only site which is left invariant under all of the symmetry operations in the site-symmetry group. We call a point which is left invariant under all of the symmetry operations a fixed point.
Note that if $\mathbf{q}_1$ and $\mathbf{q}_2$ are part of the same Wyckoff position, then their site-symmetry groups are isomorphic; thus, if $\mathbf{q}_1$ is the only fixed point of $G_{\mathbf{q}_1}$, then $\mathbf{q}_2$ is the only fixed point of $G_{\mathbf{q}_2}$. Consequently, if {any} one site in the Wyckoff position has a maximal site-symmetry group, then the site-symmetry group of any point in the Wyckoff position is maximal.

We first prove that a finite group acting on a vector space always has at least one fixed point: consider a finite group, $K = \{k_1, ..., k_n\}$.
Then, for any $i= 1,... ,n$, $\{ k_ik_1, ..., k_ik_n\} = K$ (this is evident because $k_ik_j \in K$ by group closure and for any $k_j\in K$, $k_j = k_i(k_i^{-1}k_j)$, where, again by group closure, $k_i^{-1}k_j\in K$.)
It then follows that, for an arbitrary vector $\mathbf{x}$, the sum $\sum_i k_i \mathbf{x}$ is invariant under all elements of $K$; thus, $\sum_i k_i \mathbf{x}$ is a fixed point of $K$.
This completes the proof that a finite group always has at least one fixed point.
It also implies the following useful corollary:
\begin{cor}\label{cor:nofixedinfinite}
A group which has no fixed point is infinite.
\end{cor}
{\noindent (It follows that any group containing a translation, screw, or glide symmetry is infinite.)}

We can now prove that a site-symmetry group with exactly one fixed point is maximal.
Let $G_\mathbf{q}$ be the site-symmetry group of $\mathbf{q}$ and suppose that $G_\mathbf{q}$ has only $\mathbf{q}$ as a fixed point. 
Now consider $g\in G, g\notin G_\mathbf{q}$ and define $G'_\mathbf{q}$ to be the group generated by $g$ and the generators of $G_\mathbf{q}$.
Then $G'_\mathbf{q}$ does not have any fixed points: because $g \notin G_\mathbf{q}$, $\mathbf{q}$ is not a fixed point (else $g$ would be in $G_\mathbf{q}$), but because $G_\mathbf{q}$ has no fixed points besides $\mathbf{q}$, no other point can be fixed.
Thus, $G'_\mathbf{q}$ has no fixed points and, hence, using Corollary~\ref{cor:nofixedinfinite}, $G'_\mathbf{q}$ is infinite.
It follows that there is no finite group, $H \neq G_\mathbf{q}$ such that $G_\mathbf{q} \subset H \subset G$.
Hence, according to Def~\ref{def:maximal}, $G_\mathbf{q}$ is maximal.

It follows that a non-maximal site-symmetry group leaves at least two points fixed.
Notice that if the group leaves two points fixed, then it also leaves the path containing those points fixed (i.e., if $G_\mathbf{q} \mathbf{q}_{1,2}= \mathbf{q}_{1,2}$ then $G_\mathbf{q} \left[ a_1\mathbf{q}_1 + a_2\mathbf{q}_2 \right] = a_1\mathbf{q}_1 + a_2\mathbf{q}_2$).
Similarly, if the group leaves three non-collinear points fixed, then it leaves the plane containing those points fixed.
Consequently, any non-maximal site-symmetry group leaves either a line or a plane fixed (or, in the trivial case where $G_\mathbf{q}$ only contains the identity, it leaves all of space fixed.)

We remark that it is not necessary for $\mathbf{G}_\mathbf{q}$ to have a single fixed point in order to be maximal: for example, consider SG $P6mm$, which is generated by the wallpaper group $p6mm$ and by a unit translation in the $\hat{z}$ direction.
Each Wyckoff position in $P6mm$ has the same site-symmetry group as its projection onto the $x-y$ plane.
Thus, if $G_\mathbf{q}$ leaves a single point invariant in $2D$, it leaves an entire line invariant in $3D$: for example,
$G_{1a}\cong C_{6v}$ leaves only the origin invariant in $2D$, but leaves the $\hat{z}$-axis invariant in $3D$.
This is a general feature of $3D$ SGs generated by a wallpaper group and translations in the $\hat{z}$ direction.

\section{Transformations of Wannier functions}
\label{sec:Wannierbasis}

Following Ref~\onlinecite{Evarestov1997},
{we derive how a Wannier function $W_{i\alpha}(\mathbf{r}-\mathbf{t}_\mu)$ transforms under an arbitrary element $h=\{R|\mathbf{t}\} \in G$ in the band representation $\rho_G(h)$ induced from a representation $\rho$ of $G_\mathbf{q}$, for some site $\mathbf{q}$:}
\begin{align}
\rho_G(h)W_{i\alpha}(\mathbf{r}-\mathbf{t}_\mu) &= h\{E|\mathbf{t}_\mu\} W_{i\alpha}(\mathbf{r}) \nonumber\\
&= \{E|R\mathbf{t}_\mu \} h W_{i\alpha}(\mathbf{r}) \nonumber\\
&= \{E|R\mathbf{t}_\mu \} \{E|\mathbf{t}_{\beta\alpha} \} g_\beta g g_\alpha^{-1}W_{i\alpha}(\mathbf{r}) \nonumber\\
&= \{ E| R\mathbf{t}_\mu + \mathbf{t}_{\beta\alpha} \} g_\beta g W_{i1}(\mathbf{r})\nonumber\\
&= \{ E| R\mathbf{t}_\mu + \mathbf{t}_{\beta\alpha} \} g_\beta \left[ \rho(g) \right]_{ji}W_{j1}(\mathbf{r}) \nonumber\\
&=  \{ E| R\mathbf{t}_\mu + \mathbf{t}_{\beta\alpha} \}  \left[ \rho(g) \right]_{ji} W_{j\beta}(\mathbf{r}) \nonumber\\
&=  \left[ \rho(g) \right]_{ji} W_{j\beta}(\mathbf{r}- R\mathbf{t}_\mu - \mathbf{t}_{\beta\alpha} ),
\label{eq:inducedlocal}
\end{align}
where we have used the decomposition of Eq~(\ref{eq:defbeta}); $hg_\alpha =\{E|\mathbf{t}_{\beta\alpha}\}g_{\beta}g$, for some $g\in G_\mathbf{q}$ and {coset representative $g_\beta$}; and $\mathbf{t}_{\beta\alpha}=h\mathbf{q}_\alpha-\mathbf{q}_{\beta}$ a Bravais lattice vector.

We now derive the action of $h$ on the Fourier-transformed functions $a_{i\alpha}(\mathbf{k},\mathbf{r})$, defined in Eq~(\ref{eq:fourier}):
\begin{align}
\rho_G(h) a_{i\alpha}(\mathbf{k},\mathbf{r}) &\equiv \rho_G(h) \sum_\mu e^{i\mathbf{k} \cdot \mathbf{t}_\mu} W_{i\alpha}(\mathbf{r} - \mathbf{t}_\mu) \nonumber\\
&= \sum_\mu e^{i\mathbf{k} \cdot \mathbf{t}_\mu} \left[ \rho(g) \right]_{ji} W_{j\beta}(\mathbf{r}- R\mathbf{t}_\mu - \mathbf{t}_{\beta\alpha} ) \nonumber\\
&= e^{-i(R\mathbf{k})\cdot \mathbf{t}_{\beta\alpha} } \left[ \rho(g) \right]_{ji}  \times \nonumber\\
&\quad\quad\sum_\mu e^{i(R\mathbf{k})\cdot(R\mathbf{t}_\mu + \mathbf{t}_{\beta\alpha} )} W_{j\beta}(\mathbf{r}- R\mathbf{t}_\mu - \mathbf{t}_{\beta\alpha} )  \nonumber\\
&= e^{-i(R\mathbf{k})\cdot \mathbf{t}_{\beta\alpha} } \left[ \rho(g) \right]_{ji} a_{j\beta}(R\mathbf{k}, \mathbf{r} ),
\label{eq:inducedk}
\end{align}
{where we have used the fact that $R$ is orthogonal.}
Eq~(\ref{eq:inducedk}) is exactly Eq~(\ref{eq:inducedrep}), remembering that $g$ is determined by Eq~(\ref{eq:defbeta}).

\section{{Graphene $p_z$ orbitals without inversion: example of a disconnected EBR}}
\label{sec:graphene2b}

We choose the lattice vectors of the honeycomb lattice:
\begin{align}
\mathbf{e}_1 &= \frac{\sqrt{3}}{2} \hat{\mathbf{x}} + \frac{1}{2} \hat{\mathbf{y}} \nonumber\\
\mathbf{e}_2 &= \frac{\sqrt{3}}{2} \hat{\mathbf{x}} - \frac{1}{2} \hat{\mathbf{y}},
\end{align}
which are shown in Fig~\ref{fig:graphenebasisvectors}. {Following the notation of Ref~\onlinecite{NaturePaper}, we choose the group generators}:
\begin{align}
C_{3}:& (\mathbf{e}_1,\mathbf{e}_2)\rightarrow(-\mathbf{e}_2,\mathbf{e}_1-\mathbf{e}_2)\label{eq:grpaction1}\\
C_{2}:& (\mathbf{e}_1,\mathbf{e}_2)\rightarrow(-\mathbf{e}_1,-\mathbf{e}_2)\\
m_{1\bar{1}}:&(\mathbf{e}_1,\mathbf{e}_2)\rightarrow(\mathbf{e}_2,\mathbf{e}_1);\label{eq:grpaction2}
\end{align}
the subscript $1\bar{1}$ denotes that the mirror line has normal vector $\mathbf{e}_1-\mathbf{e}_2= \hat{\mathbf{y}}$.


We consider spinful $p_z$ orbitals on the corners of the honeycombs (the $2b$ position in Fig~\ref{fig:grapheneWyckoff}), as in graphene.
We define the sites $\mathbf{q} \equiv \mathbf{q}_1 \equiv (\frac{1}{3}, \frac{1}{3})$ and $ \mathbf{q}_2 \equiv (-\frac{1}{3},-\frac{1}{3})$.
The site symmetry group $G_\mathbf{q}$ is generated by $\{C_{3}|01\}$ and $\{m_{1\bar{1}}|00\}$; the group is isomorphic to $C_{3v}$.
We choose the matrix representation:
\begin{align}
\rho(\{C_{3z}|01\}) &= e^{\frac{i \pi}{3}s_z}\nonumber\\ 
\rho(\{m_{1\bar{1}}| \mathbf{0} \}) &= is_x,
\label{eq:C3vrep}
\end{align}
where $s_{x,y,z}$ are the Pauli matrices. 
This choice for the representative of $m_{1\bar{1}}$ differs by a unitary transformation from the basis where $m_{1\bar{1}}$ is represented by $e^{i\pi s_y/2} = is_y$ (which is the natural basis for a $\pi$ spin rotation about the $\mathbf{e}_1-\mathbf{e}_2=\hat{\mathbf{y}}$ axis); we choose it here to be consistent with Ref~\onlinecite{NaturePaper} and \onlinecite{Kane05}.
Comparing the characters with Table~\ref{table:c3v} shows that the representation in Eq~(\ref{eq:C3vrep}) is the spin-$\frac{1}{2}$ representation, $\bar{\Gamma}_6$.

\begin{table}[h]
{\renewcommand{\arraystretch}{1.2}%
\begin{tabular}{c|cccc}
Rep & $E$ & $C_{3}$ & $m$ & $\bar{E}$ \\
\hline 
$\bar{\Gamma}_4$ & 1 & -1 & -i & -1 \\ 
$\bar{\Gamma}_5$ & 1 & -1 & i & -1 \\
$\bar{\Gamma}_6$ & 2 & 1 & 0 & -2
\end{tabular}
}
\caption{Character table for the double-valued representations of $C_{3v}$.\cite{GroupTheoryPaper} 
The one-dimensional representations $\bar{\Gamma}_4$ and $\bar{\Gamma}_5$ are complex conjugates of each other. The two dimensional $\bar{\Gamma}_{6}$ representation is the spin-$\half$ representation, while the one-dimensional $\bar{\Gamma}_4$ and $\bar{\Gamma}_5$ representations act in the space of spin $|S=3/2, m_z=3/2\rangle\pm i|S=3/2,m_z=-3/2\rangle$ respectively.}\label{table:c3v}
\end{table}

\subsection{Characters of EBRs at high-symmetry points}
\label{sec:charactersgraphene}

We want to compute the characters {at high-symmetry points, $\mathbf{k}$}, of the band representation induced from $p_z$ orbitals {on the $2b$ Wyckoff position}.
For pedagogical purposes, we {explicitly} construct the matrix representatives here using Eq~(\ref{eq:inducedrep}), instead of skipping to the character formula in Eq~(\ref{eq:character}).
The matrix representatives were computed in Ref~\onlinecite{NaturePaper} (see Eqs (S21)--(S25)) using an intuitive constructive that differs by a unitary transformation from Eq~(\ref{eq:inducedrep}) in this paper.

We first choose the coset representatives: $g_1 = \{E| \mathbf{0}\}, g_2 =\{ C_{2} | \mathbf{0} \}$, which satisfy $g_\alpha \mathbf{q} = \mathbf{q}_\alpha$.
The next step is to evaluate Eq~(\ref{eq:defbeta}), which we rewrite here for convenience:
\begin{equation*}
hg_\alpha =\{E|\mathbf{t}_{\beta\alpha}\}g_{\beta}g  \tag{\ref{eq:defbeta}}
\end{equation*}
  
We can then compute the band representation matrices using Eq~(\ref{eq:inducedrep}).
Given $h$ and $g_\alpha \in \{ g_1, g_2\}$ on the left-hand side of Eq~(\ref{eq:defbeta}), we need find the lattice translation $\mathbf{t}_{\beta\alpha}$, $g_\beta\in \{g_1,g_2\}$ and $g\in G_\mathbf{q}$ that satisfy Eq~(\ref{eq:defbeta}).
{We now do that} for each generator, $h$, of the honeycomb lattice:
\begin{description}
\item[$h=\{ C_{3}| \mathbf{0} \} $]
In this case, Eq~(\ref{eq:defbeta}) is written as:
\begin{align} 
\{ C_{3}| \mathbf{0} \} g_1 &= \{E | 0 \bar{1} \} g_1 \{C_{3}|01\} \nonumber\\
\{ C_{3}| \mathbf{0} \} g_2 &= \{E | 0 1 \} g_2 \{C_{3}|01\}
\label{eq:defbetaC3z}
\end{align}
Because $\{ C_{3} | \mathbf{0} \}$ does not mix the two sites in the Wyckoff position (instead, it shifts $\mathbf{q}_{1,2}$ by lattice vectors), $\alpha = \beta$ in both lines of Eq~(\ref{eq:defbetaC3z}); this will be true for any {$h$} which differs from an element of $G_\mathbf{q}$ by a lattice translation.

{To find $(\rho \uparrow G)\downarrow G_\mathbf{k}$} when $\{ C_{3}| \mathbf{0} \} \in G_\mathbf{k}$ (recall, $G_\mathbf{k}$ consists of all space group operations which leave $\mathbf{k}$ unchanged up to a reciprocal lattice vector), 
we apply Eq~(\ref{eq:inducedrep}) to Eq~(\ref{eq:defbetaC3z}), which yields:
\begin{equation}
\rho_G^\mathbf{k}(\{ C_{3}| \mathbf{0} \}) = \! \begin{pmatrix} e^{i\mathbf{k}\cdot \mathbf{e}_2 } & 0 \\ 0 & e^{-i\mathbf{k}\cdot \mathbf{e}_2} \end{pmatrix} \! \otimes \!  e^{\frac{i\pi }{3}s_z }
\label{eq:bandrepC3z}
\end{equation}

\item[$h=\{ m_{1\bar{1}} | \mathbf{0} \}$]
In this case, Eq~(\ref{eq:defbeta}) yields
\begin{align} 
\{ m_{1\bar{1}} | \mathbf{0} \} g_1 &= \{E | \mathbf{0} \} g_1 \{ m_{1\bar{1}}| \mathbf{0} \}\nonumber\\
\{ m_{1\bar{1}} | \mathbf{0} \} g_2 &= \{E | \mathbf{0} \} g_2 \{ \bar{m}_{1\bar{1}}| \mathbf{0} \},
\label{eq:defbetam11b}
\end{align}
where the $\bar{m}_{1\bar{1}}$ denotes the combined operation $m_{1\bar{1}}$ followed by a $2\pi$ {spin} rotation; notice that the $2\pi$ rotation is the product of two consecutive operations of $\{ m_{1\bar{1}}| \mathbf{0} \}$, which makes it an element of $G_\mathbf{q}$. It is nontrivial because it imparts an overall minus sign in our double-valued (spinful) representation.
Thus,
\begin{equation}
\rho_G^\mathbf{k}(\{ m_{1\bar{1}} | \mathbf{0} \}) = \! \begin{pmatrix} is_x & 0 \\ 0 & -is_x\end{pmatrix} \! = \! \sigma_z \! \otimes is_x
\label{eq:bandrepm11b}
\end{equation}
when $\{ m_{1\bar{1}} | \mathbf{0} \} \in G_\mathbf{k}$.

\item[$h=\{ C_{2} |  \mathbf{0}\}$]
In this case, Eq~(\ref{eq:defbeta}) yields:
\begin{align} 
\{ C_{2}| \mathbf{0} \} g_1 &= \{E |  \mathbf{0} \} g_2 \{E |  \mathbf{0}\} \nonumber\\
\{ C_{2}| \mathbf{0} \} g_2 &= \{E |  \mathbf{0}\} g_1 \{\bar{E} | \mathbf{0}\},
\label{eq:defbetaC2z}
\end{align}
which yields the subduced representation when $\{C_{2}| \mathbf{0}\}$ is in the little group of $\mathbf{k}$:
\begin{equation}
\rho_G^\mathbf{k}(\{C_{2} |  \mathbf{0}\}) = \begin{pmatrix}  0 & -\mathbb{I} \\ \mathbb{I} & 0 \end{pmatrix} = -i\sigma_y\otimes \sigma_0,
\label{eq:bandrepC2z}
\end{equation}
where we have {again} used the fact that $\rho(\{E|\mathbf{0}\}) = - \rho(\{ \bar{E} | \mathbf{0} \})= \mathbb{I}$.
\end{description}

%
\begin{figure}[h]
	\centering
	\includegraphics[width=1in]{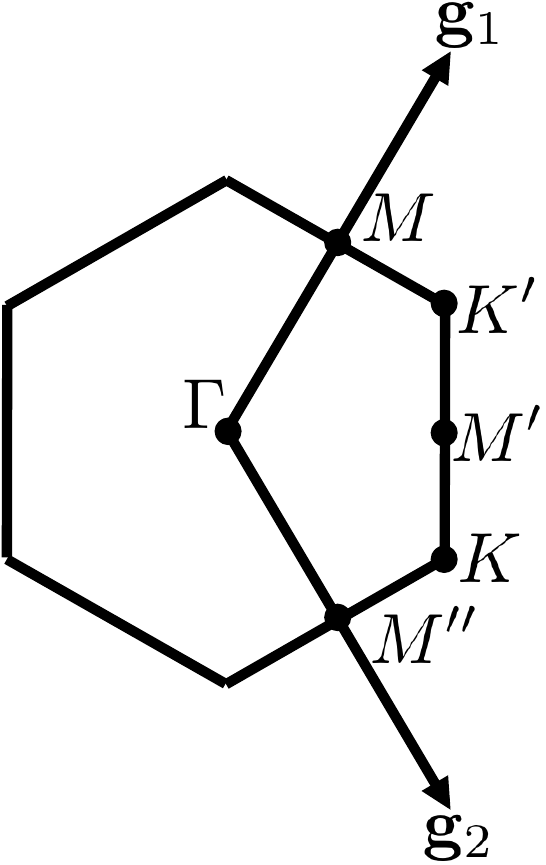}
	\caption{Reciprocal lattice basis vectors and high-symmetry points of the hexagonal lattice.}
	\label{fig:graphenereciprocal}
\end{figure}

 \begin{table}[h]
 {\renewcommand{\arraystretch}{1.2}%
\begin{tabular}{cc}
$\mathbf{k}$ & Irreps \\
 \hline
$\Gamma$ & $\bar{\Gamma}_8 \oplus \bar{\Gamma}_9$\\
$K$ & $\bar{K}_4 \oplus \bar{K}_5 \oplus \bar{K}_6$\\
$M$ & $\bar{M}_5 \oplus \bar{M}_5$
\end{tabular}
}
\caption{Irreps of the little groups that appear at high-symmetry points -- $\Gamma$, $K$ and $M$ -- in graphene, labelled by the irreps of the corresponding little co-groups, which are isomorphic to $C_{6v}$, $C_{3v}$ and $C_{2v}$, respectively.
The character tables for these groups are given in Tables~\ref{table:c6v}, \ref{table:c3v}, and \ref{table:c2v}.
While all irreps of the abstract point groups are denoted by $\bar{\Gamma}_n$, we label the irreps at $K$ and $M$ by $\bar{K}_n$ or $\bar{M}_n$.}
\label{tab:grapheneirreps}
\end{table}

We will now compute the characters of the matrix representations in Eqs~(\ref{eq:bandrepC3z}), (\ref{eq:bandrepm11b}) and (\ref{eq:bandrepC2z}) at the high-symmetry points $\Gamma = (0,0),  K =(\frac{1}{3},\frac{2}{3}), M=(\frac{1}{2},0)$, defined with respect to the reciprocal lattice vectors shown in Fig~\ref{fig:graphenereciprocal}.
We compare to {the characters in} Tables~\ref{table:c6v}, \ref{table:c3v}, and \ref{table:c2v} to determine the multiplicity of each little group irrep in the appropriate subduced representation; this {method} corresponds to Eq~(\ref{eq:decompose}) in the main text.
The results are shown in Table~\ref{tab:grapheneirreps}.

The little co-group at $\Gamma$ is isomorphic to $C_{6v}$. We know $\chi_G^\Gamma(\{ E | \mathbf{0} \} )=4$ and $\chi_G^\Gamma(\{ C_{3} | \mathbf{0} \} )=2$ (per Eq~(\ref{eq:bandrepC3z})). Since $\{ C_{6} | \mathbf{0} \} \notin G_\mathbf{q}$, we also know that $\chi_G^\Gamma(\{ C_{6} | \mathbf{0} \} )=0$. Comparison to Table~\ref{table:c6v} shows that $\rho_G^\Gamma = \bar{\Gamma}_8 \oplus \bar{\Gamma}_9$ {($\bar{\Gamma}_{8,9}$ are both $2D$ irreps.)}

The little co-group at $K$ is generated by $C_{3}$ and $C_{2}m_{1\bar{1}}$. 
We know $\chi_G^K(\{ E | \mathbf{0} \} )=4$.
Per Eq~(\ref{eq:bandrepC3z}), $\chi_G^K(\{ C_{3} | \mathbf{0} \} )=-1$. Further, since $\{ C_{2}m_{1\bar{1}}| \mathbf{0} \} \notin G_\mathbf{q}$ {(which follows from $\{m_{1\bar{1}}|\mathbf{0}\}\in G_\mathbf{q}$, $\{ C_{2}| \mathbf{0} \} \notin G_\mathbf{q}$)}, $\chi_G^K( \{ C_{2}m_{1\bar{1}}| \mathbf{0} \} ) = 0$. 
Comparison to Table~\ref{table:c3v} shows that $\rho_G^K = \bar{K}_4 \oplus \bar{K}_5 \oplus \bar{K}_6$ (notice that {in character tables} we use the notation $\bar{\Gamma}_{4,5,6}$ to refer to the irreps of an abstract group -- in this case $C_{3v}$ -- but the notation $\bar{K}_{4,5,6}$ to refer to the irreps of the little group at $K$.)

Finally, since the little co-group of $M$ is isomorphic to $C_{2v}$, which has only one double-valued irrep, as shown in Table~\ref{table:c2v}, the fact that $\rho_G^M$ is four-dimensional is enough to tell that $\rho_G^M = 2\bar{M}_5$ (again, we use $\bar{\Gamma}_5$ to refer to the irrep of the abstract group $C_{2v}$, but the notation $\bar{M}_5$ to refer to the irrep of the little group at $M$.)

 \begin{table}[h]
 {\renewcommand{\arraystretch}{1.2}%
\begin{tabular}{c|ccccc}
 Rep & $E$ & $C_{2}$ & $m$ & $C_{2}m$ &$\bar{E}$ \\
 \hline
$\bar{\Gamma}_5$ & 2 & 0 & 0 & 0 & -2
\end{tabular}
}
\caption{Character table for the double-valued irrep of $C_{2v}$.\cite{GroupTheoryPaper} $\bar{\Gamma}_5$, is the two-dimensional spin-$\half$ representation.  In terms of the Pauli matrices, it is given concretely as $\bar{\Gamma}_5(C_{2})=i\sigma_z,\bar{\Gamma}_5(m)=i\sigma_y$.}
\label{table:c2v}
\end{table}

While we omit it here for brevity, we can repeat the induction procedure for any irrep of the site-symmetry group of any Wyckoff position and then subduce to the little groups of the high-symmetry points in the Brillouin zone.
The results for the {double-valued irreps of the other} maximal Wyckoff positions of the honeycomb lattice are shown in Table~\ref{table:gammareps}.

%

\begin{table}[h]
 {\renewcommand{\arraystretch}{1.2}%
\begin{tabular}{c|c|c|c|c|c}
WP & $\rho$ & $\Gamma$ & $K$ & $M$ & d \\
\hline
$1a(C_{6v})$ & $\bar{\Gamma}_7$ & $\bar{\Gamma}_7$ & $\bar{K}_4\oplus\bar{K}_5$ & $\bar{M}_5$ & 2 \\
& $\bar{\Gamma}_8$ & $\bar{\Gamma}_8$ & $\bar{K}_6$ & $\bar{M}_5$ & 2 \\
& $\bar{\Gamma}_9$ & $\bar{\Gamma}_9$ & $\bar{K}_6$ & $\bar{M}_5$ & 2 \\
$2b(C_{3v})$ & $\bar{\Gamma}_4$ & $\bar{\Gamma}_7$ & $\bar{K}_6$ & $\bar{M}_5$ & 2 \\
 & $\bar{\Gamma}_5$ & $\bar{\Gamma}_7$ & $\bar{K}_6$ & $\bar{M}_5$ & 2 \\
 & $\bar{\Gamma}_6$ & $\bar{\Gamma}_8\oplus\bar{\Gamma}_9$ &$\bar{K}_4\oplus\bar{K}_5\oplus\bar{K}_6$ & $2\bar{M}_5 $& 4 \\
$3c(C_{2v})$ & $\bar{\Gamma}_5$ & $\bar{\Gamma}_7\oplus \bar{\Gamma}_8 \oplus \bar{\Gamma}_9$ & $\bar{K}_4\oplus \bar{K}_5\oplus 2\bar{K}_6 $ & $3\bar{M}_5$ & 6
\end{tabular}
}
\caption{{EBRs} induced from double-valued irreps\cite{GroupTheoryPaper} of the maximal site-symmetry groups in $p6mm$. As explained in Appendix~\ref{sec:exceptionexample}, the band representation induced from the $3c$ position is composite -- it furnishes an exception -- unless time reversal symmetry is present. 
The first column lists a maximal Wyckoff position and the point group isomorphic to its site-symmetry group.
The second column gives the irrep of the maximal site-symmetry group, $G_\mathbf{q}$, from which the band representation is induced. 
The third column gives the little group representations which appear in the induced EBR at the $\Gamma$ point, as defined in Table~\ref{table:c6v}. The fourth column gives the little group irreps that appear at the $K$ point, as defined in Table~\ref{table:c3v}. The fifth column gives the little group irreps that appear at the $M$ point, as defined in Table~\ref{table:c2v}. The last column gives the dimension of the EBR, which is also the connectivity of the elementary band rep. }\label{table:gammareps}
\end{table}

\subsection{Connectivity of the EBRs}

\begin{table}[h]
 {\renewcommand{\arraystretch}{1.2}%
\begin{tabular}{c|ccccc}
 Rep & $E$ & $m$ & $\bar{E}$ \\
 \hline
$\bar{\Gamma}_3$ & $1$ & $-i$ & $-1$\\
$\bar{\Gamma}_4$ & $1$ & $i$ & $-1$
\end{tabular}
}
\caption{Character table for the double-valued irreps of $C_s$.\cite{GroupTheoryPaper}}
\label{table:cs}
\end{table}

We want to know whether the EBR induced from $p_z$ orbitals on the honeycomb lattice (derived in the previous section) is connected.
To this end, we derive the compatibility relations introduced in Sec~\ref{sec:compatibility}.
At each high-symmetry point ($\Gamma, K, M$), we will decompose the little group irreps that appear at that point into a sum of irreps of the little group of each high-symmetry line emanating from that point.
There are three high-symmetry lines:\cite{GroupTheoryPaper,NaturePaper}: $\Gamma-K$, $\Gamma-M$ and $K-M$.
Although the little group of each line is distinct, all three groups are isomorphic to $C_s\ltimes\mathbb{Z}^2$, generated by a single mirror and two primitive lattice translations.
Table~\ref{table:cs} provides the character table of $C_s$.
By comparing to the character tables of $C_{6v}, C_{3v}$ and $C_{2v}$, we see that the
{two-dimensional irreps of $C_{6v}, C_{3v}$ and $C_{2v}$ always subduce to $\bar{\Gamma}_3\oplus\bar{\Gamma}_4$, while the one-dimensional irreps, $\bar{\Gamma}_{4,5}$, of $C_{3v}$ subduce to the one-dimensional irreps $\bar{\Gamma}_{3,4}$, respectively, of $C_s$.}



We now consider the band representations in Table~\ref{table:gammareps}; the last column of the table gives the dimension of the band representation.
Since $\bar{M}_5$ is two-dimensional (c.f. Table~\ref{table:c2v}), any band representation that is two-dimensional must be connected, since its bands at least connect at $\bar{M}_5$.
Hence, the first five EBRs listed in Table~\ref{table:gammareps} are connected.

\begin{figure}[h]
\centering
\subfloat[]{
	\includegraphics[width=3in]{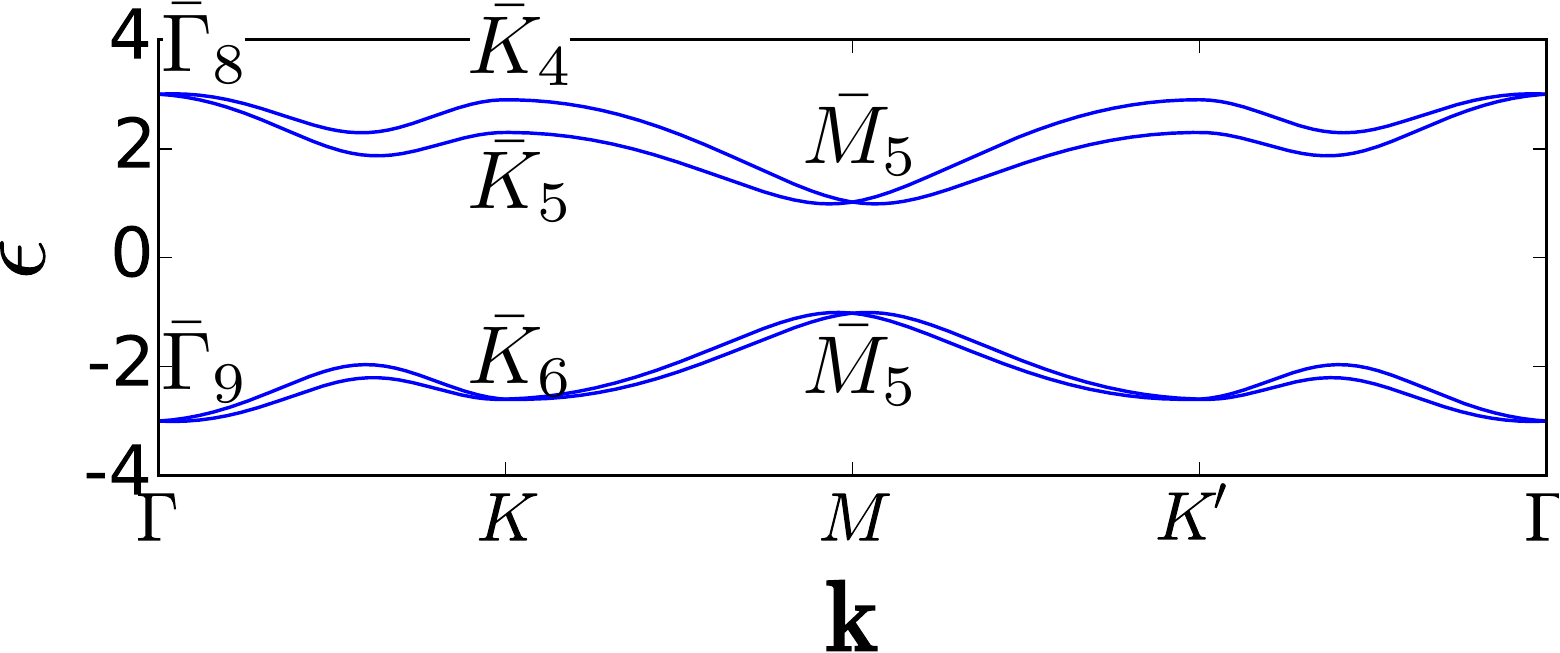}
	\label{fig:graphenedisconnected}
}
\hspace{.1in}
\subfloat[]{
	\includegraphics[width=3in]{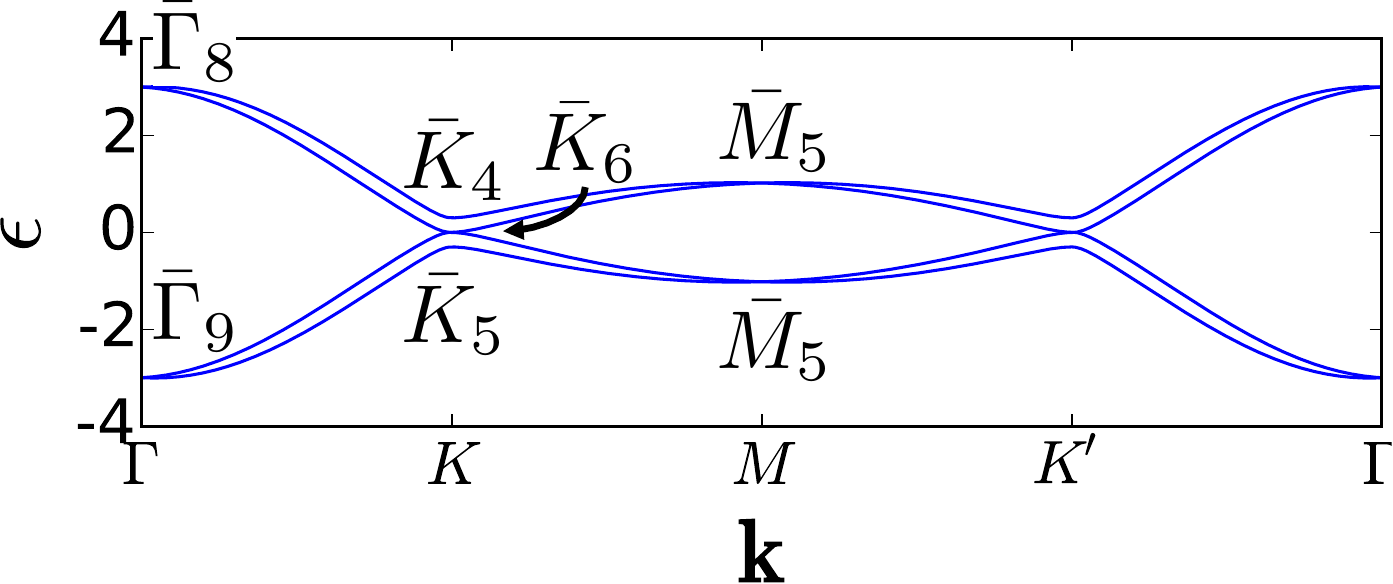}
	\label{fig:grapheneconnected}
}
\caption{Two possible connectivities for the EBR induced from the $2b$ position, $\bar{\Gamma}_6\uparrow G$ with little group irreps labelled. In (a) the bands are disconnected {and thus the band structure contains a group of topological bands}, while in (b), they are connected. Energetics determine which phase occurs in a particular system.}
\label{fig:grapheneconnectivity}
\end{figure}

However, as we now show, the compatibility relations allow for the band representation induced from $\bar{\Gamma}_6$ on the $2b$ position (as well as from $\bar{\Gamma}_5$ on the $3c$ position) to be disconnected.\cite{NaturePaper,GraphTheoryPaper,GraphDataPaper}
In this case, the bands can split into two disconnected components, the first consisting of $\bar{\Gamma}_8, \bar{K}_4, \bar{K}_5, $ and $\bar{M}_5$ and the second consisting of $\bar{\Gamma}_9, \bar{K}_6$ and $\bar{M}_5$.
The situation is depicted in Fig.~\ref{fig:graphenedisconnected}. 
{According to Proposition~\ref{prop:connectivity}, at least one of the groups of bands that comprise the disconnected EBR is topological.}

The topological bands are protected by $C_{2}$ symmetry, which can be deduced by checking on the BCS server.\cite{GroupTheoryPaper,progbandrep}
The space group $R3m$ (SG 160), generated by $C_{3z}$ and $m_{1\bar{1}}$, which is a subgroup of the space group $P6mm$ (SG 183) that describes layers of graphene, does not have any disconnected EBRs without time reversal symmetry. Thus, if $C_{2z}$ symmetry is removed, then the bands are topologically trivial. One can also check that the topological protection is independent of the mirror symmetry, since the band representation induced in space group $P6$ (SG 168), generated by $C_{3z}$ and $C_{2z}$, from the $2b$ position {also} yields a disconnected EBR {identical to that in $P6mm$.}

It is also possible for the bands to be connected, as depicted in Fig~\ref{fig:grapheneconnected}.
Whether the bands are connected or disconnected depends on energetics and the strength of the different spin orbit coupling terms present in the sample.
A similar situation is true for the band representation induced from $\bar{\Gamma}_5$ on the $3c$ position.

\subsection{{Comparison to eigenvalue classification}}
\label{sec:comparison}

{We now specify to the Kane-Mele model of graphene\cite{Kane05} with sublattice symmetry, time-reversal symmetry, and inversion symmetry-breaking Rashba SOC:
\begin{equation}
H = t_i \sum_{\langle ij\rangle}c_i^\dagger c_j + i\lambda_{SO}\sum_{\langle\langle ij\rangle\rangle }\nu_{ij}c_i^\dagger s^zc_j + i\lambda_R \sum_{\langle ij\rangle} c_i^\dagger (\mathbf{s}\times \hat{\mathbf{d}}_{ij})_zc_j
\label{eq:kanemele}
\end{equation}
Since this model is derived from $p_z$ orbitals on the $2b$ position, its bands are described by the EBR $\bar{\Gamma}_6 \uparrow G$.
The band structure is shown in Fig~\ref{fig:graphenedisconnected} in the $\mathbb{Z}_2$ nontrivial phase. }

It is immediately evident that the two connected upper bands (which contain the irreps $\bar{\Gamma}_8, \bar{K}_4, \bar{K}_5, $ and $\bar{M}_5$) are topological because they cannot be decomposed into any of the EBRs {in $P6mm$}, given in Table~\ref{table:gammareps}.
In contrast, the irreps that appear in the lower bands {($\bar{\Gamma}_9, \bar{K}_6$ and $\bar{M}_5$)} are identical to the irreps that appear in the EBR induced from the $1a$ position, $\bar{\Gamma}_9 \uparrow G$, as can also be seen in Table~\ref{table:gammareps}.
Both sets of bands have a nontrivial $\mathbb{Z}_2$ invariant, even though the irreps that appear in the lower bands match those in $\bar{\Gamma}_9 \uparrow G$.
This example was also given in Ref~\onlinecite{NaturePaper} as an example of a disconnected EBR that yields topological bands, although it was not directly stated in that paper that we were referring to the Kane-Mele model.
(It is also possible to construct a different Hamiltonian where the bands exhibit the same irreps, but have a trivial $\mathbb{Z}_2$ index.\cite{PoFragile})

Recent works\cite{Kruthoff2016,Po2017} have classified noninteracting fermionic phases by using a vector, $\mathbf{v}$, to describe a group of bands, where each component of the vector indicates the number of times a particular irrep of a particular high-symmetry point appears in that group of bands. 
For example, the number of times the $\Gamma_j$ irrep appears in the subduction of the band representation at $\Gamma$ would be denoted by $v_{\Gamma,j}$.
The compatibility relations restrict the set of allowed vectors.
Heuristically, these classification schemes consider the set of vectors that satisfy the compatibility relations for a particular space group modulo a set of ``trivial'' vectors, where a trivial vector is one which can be obtained from an atomic limit.
Notice that this classification scheme is contained within our theory of elementary band representations, since the elementary band representations define the irreps that appear at each high-symmetry point in the Brillouin zone.

Such a classification will assign the lower bands in Fig.~\ref{fig:graphenedisconnected} a trivial index, even though the bands are topological, because their irreps match those of $\bar{\Gamma}_9 \uparrow G$.
In addition, the upper bands will also be assigned a trivial index because their irreps can be obtained from a subtraction of the irreps in $\bar{\Gamma}_9 \uparrow G$ from the irreps in $\bar{\Gamma}_6 \uparrow G$, even though the irreps in the upper bands cannot be decomposed into any of the EBRs in $P6mm$, given in Table~\ref{table:gammareps}.
This is consistent with the discussion and observation in Ref.~\onlinecite{Po2017}.


\section{Example of two EBRs that share the same irreps at each $\mathbf{k}$ point but are not equivalent}
\label{sec:ExNotEquiv}

In this section we explicitly work out an example of two EBRs that at each point, $\mathbf{k}$, decompose into the same irreps of $G_\mathbf{k}$, but which differ by a Berry phase; this example has been examined in Refs~\onlinecite{Bacry1988b,Bacry1993,Michel1992}, but we write it here in modern notation.
This example motivates the need for a stronger definition of equivalence than comparing the irreps of $G_\mathbf{k}$. Def~\ref{def:equiv} ensures that equivalent EBRs share all the same Wilson loop invariants.

We consider the space group $F222$ (SG 22), which describes a face-centered cubic lattice whose symmetries are generated by $\{ C_{2x}| \mathbf{0} \}, \{ C_{2y}| \mathbf{0} \}, \{ C_{2z}| \mathbf{0} \}$. 
We define the primitive unit cell lattice vectors, $\mathbf{e}_i$, and reciprocal lattice vectors, $\mathbf{g}_i$, by:
\begin{align}
\mathbf{e}_1 &= \frac{1}{2}(1,1,0), & \mathbf{g}_1&= 2\pi(1,1,-1) \nonumber\\
\mathbf{e}_2 &= \frac{1}{2}(1,0,1), & \mathbf{g}_2 &= 2\pi(1,-1,1) \nonumber\\
\mathbf{e}_3 &= \frac{1}{2}(0,1,1), & \mathbf{g}_3&=2\pi(-1,1,1)
\label{eq:vectors222}
\end{align}

\begin{figure}[h]
\centering
\subfloat[]{
	\includegraphics[width=2in]{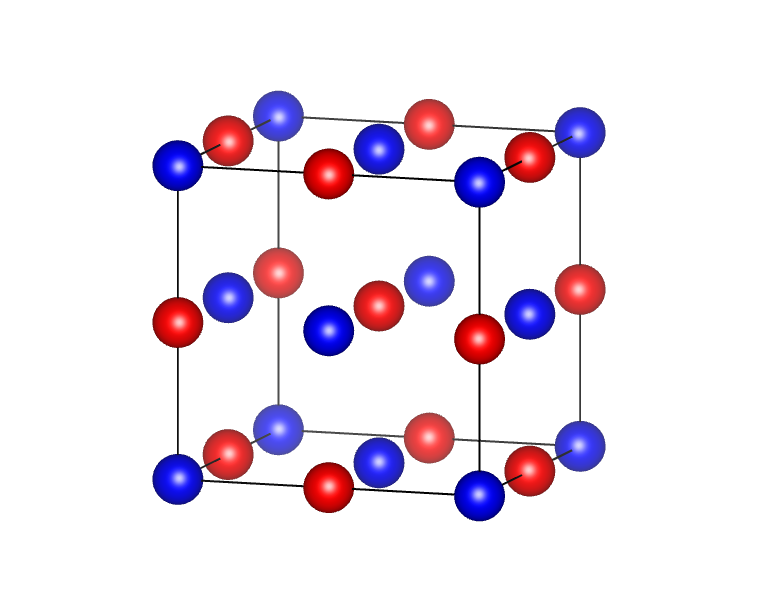}
	\label{fig:SG22unit}
	} 
\subfloat[]{
   \includegraphics[height=1.5in]{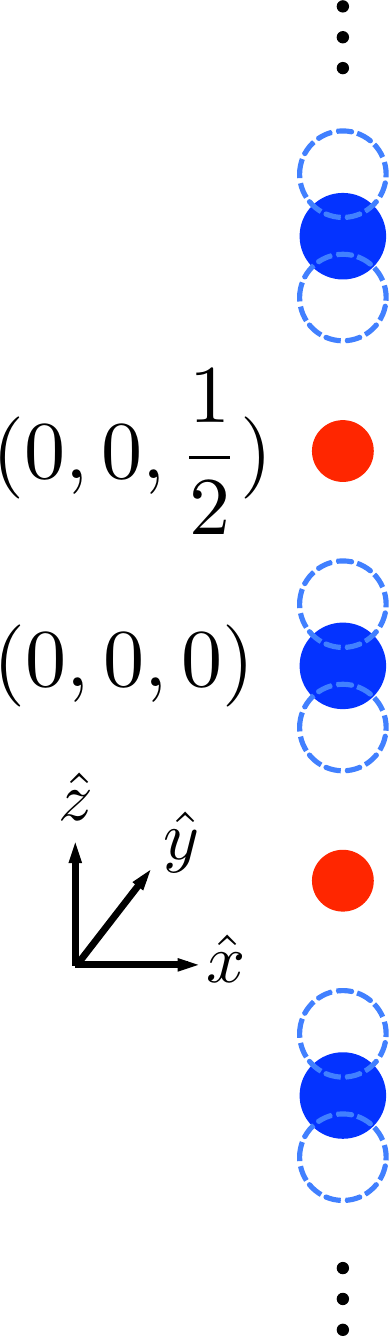}
   \label{fig:SG22horiz}
   }
\caption{{(a) Conventional unit cell for $F222$ (SG 22). The solid blue atoms sit at the $4a$ position, $(0,0,0)$, and the solid red atoms at the $4b$ position, $(0,0,1/2)$.}
(b) {Examining the $\hat{z}$ axis} shows that it is impossible to move the blue atoms to the positions of the red atoms while continuously preserving $C_{2x}$ symmetry because the symmetry requires them to move in pairs (depicted by the blue dashed circles) that transform into each other under $C_{2x}$. However, there are not enough atoms for this symmetry-preserving process to occur {(or, equivalently, the dimensionality of the site-symmetry group irrep is not big enough)}.}
\label{fig:SG22}
\end{figure}

We consider the sites $\mathbf{q}=(0,0,0)$ and $\mathbf{q}'=(0,0,1/2)$, which are described by the $4a$ and $4b$ Wyckoff positions, respectively.
Each Wyckoff position contains four sites in the conventional unit cell,
{depicted in Fig~\ref{fig:SG22unit}}, and one site in the primitive unit cell.
Their site-symmetry groups are $G_\mathbf{q} = \{ \{ C_{2x}| \mathbf{0} \}, \{ C_{2y}| \mathbf{0} \}, \{ C_{2z}| \mathbf{0} \} \}$ and $G_{\mathbf{q}'}= \{ \{C_{2x}| \mathbf{t}_z \}, \{C_{2y}| \mathbf{t}_z \}, \{ C_{2z}| \mathbf{0} \} \} $, where $\mathbf{t}_z$ indicates a translation by $-\mathbf{e}_1 + \mathbf{e}_2 + \mathbf{e}_3 = (0,0,1)$. 
The two site-symmetry groups are isomorphic to each other. 

We take $\rho$ to be the trivial representation of $G_\mathbf{q}$ and $\rho'$ the trivial representation of $G_{\mathbf{q}'}$. This corresponds physically to, for example, spinless $s$ orbitals on the relevant Wyckoff position.
Inducing the band representations $\rho_G, \rho'_G$ according to Eq~(\ref{eq:inducedrep}) is simplified by the fact that the indices $i$ and $\alpha$ are trivial ($i$ is trivial because there is one orbital on the site and $\alpha$ because there is one site in the Wyckoff position).
Thus, given $h=\{R|\mathbf{t}_h\}$, an arbitrary element in SG $F222$, Eq~(\ref{eq:inducedrep}) simplifies to 
\begin{equation}
\rho_G^\mathbf{k}(h) = e^{-i(R\mathbf{k})\cdot \mathbf{t}_h },
\end{equation}
where we have used the decomposition in Eq~(\ref{eq:defbeta}) with $\alpha = \beta = 1$, which simplifies to $h=\{E|\mathbf{t}_h\}\{ R| \mathbf{0} \}$, since all rotations $\{ C_{2x,2y,2z}| \mathbf{0} \}$ are in $G_\mathbf{q}$.
Similarly,
\begin{equation}
(\rho')_G^\mathbf{k}(h) = e^{-i(R\mathbf{k})\cdot \mathbf{t}'_h} 
\end{equation}
where
\begin{equation} 
\mathbf{t}'_h = 
\begin{cases} \mathbf{t}_h & \text{if } R= C_{2z}\\ \mathbf{t}_h-\mathbf{t}_z & \text{if }R=C_{2x,2y}
\end{cases}
\label{eq:deftprime}
\end{equation}
Eq~(\ref{eq:deftprime}) follows from the decomposition in Eq~(\ref{eq:defbeta}) with $\alpha=\beta=1$; explicitly,
if $R=C_{2z}$ then $\{ C_{2z}|\mathbf{t}_h\} = \{E|\mathbf{t}_h\} \{ C_{2z}| \mathbf{0} \}$ (notice $\{ C_{2z}| \mathbf{0} \} \in G_{\mathbf{q}'}$) and if
$R=C_{2x,2y}$ then $\{C_{2x,2y}|\mathbf{t}_h\} = \{E| \mathbf{t}_h - \mathbf{t}_z\} \{C_{2x,2y}|\mathbf{t}_z \}$ (notice $\{C_{2x,2y}|\mathbf{t}_z \}\in G_{\mathbf{q}'}$).

We now prove that the characters defined in Eq~(\ref{eq:character}),
\begin{equation}
\chi_G^\mathbf{k}(h)= \begin{cases} e^{-i\mathbf{k}\cdot \mathbf{t}_h} & \text{if }h\in G_\mathbf{k}\\
0 & \text{else}
\end{cases}
\label{eq:character222-1}
\end{equation}
and
\begin{equation}
(\chi')_G^\mathbf{k}(h)= \begin{cases} e^{-i\mathbf{k}\cdot \mathbf{t}'_h} & \text{if }h\in G_\mathbf{k}\\
0 & \text{else},
\end{cases}
\label{eq:character222-2}
\end{equation}
are equal for all $\mathbf{k}$ and $h$. 
{We have used the fact that since $R\in G_\mathbf{k}$ and $\mathbf{t}_h$ is a lattice vector, $e^{-i(R\mathbf{k})\cdot \mathbf{t}_h} = e^{-i\mathbf{k}\cdot \mathbf{t}_h}$; the same holds for $\mathbf{t}_h'$.}
Eqs~(\ref{eq:character222-1}) and (\ref{eq:character222-2}) show that $\chi_G^\mathbf{k}(h)=(\chi')_G^\mathbf{k}(h)$ when either $R=C_{2z}$ (in which case $\mathbf{t}_h = \mathbf{t}'_h$) and/or $R\notin G_\mathbf{k}$ {(in which case both characters are zero.)}

It remains to show that $\chi_G^\mathbf{k}(h)=(\chi')_G^\mathbf{k}(h)$ when $R=C_{2x,2y}$ and $R\in G_\mathbf{k}$.
Lets take $R=C_{2x}$; an analogous proof applies when $R=C_{2y}$.
The condition $R\in G_\mathbf{k}$ implies $C_{2x}\mathbf{k} = \mathbf{k}$ modulo a reciprocal lattice vector.
Utilizing Eq~(\ref{eq:vectors222}), $C_{2x}\mathbf{g}_{1,2} = \mathbf{g}_{2,1}$ and $C_{2x}\mathbf{g}_3 = -(\mathbf{g}_1+\mathbf{g}_2 + \mathbf{g}_3)$.
Thus, writing $\mathbf{k} = k_i\mathbf{g}_i$, where $k_{1,2,3}$ are defined mod $\mathbb{Z}$, the condition $h\in G_\mathbf{k}$ requires
\begin{equation}
{k_1 =k_2-k_3, \,\, k_2 = k_1 - k_3} 
\label{eq:enforcelittlegroup}
\end{equation}
Thus, $\mathbf{k}\cdot \mathbf{t}_z = k_i (\mathbf{g}_i\cdot \mathbf{t}_z) = 2\pi(-k_1 + k_2 + k_3) = 0\mod 2\pi$, where the last equality follows from Eq~(\ref{eq:enforcelittlegroup}).
Consequently, when $R=C_{2x}$, $e^{-i\mathbf{k}\cdot \mathbf{t}_h }= e^{-i\mathbf{k}\cdot \mathbf{t}'_h}$ and $\chi_G^\mathbf{k}(h)=(\chi')_G^\mathbf{k}(h)$.

We have shown that for all $h$ and $\mathbf{k}$, $\chi_G^\mathbf{k}(h)=(\chi')_G^\mathbf{k}(h)$.
It follows that the band representations $\rho_G$ and $\rho'_G$ share the decomposition into irreps of the little group $G_\mathbf{k}$ at all points $\mathbf{k}$ in the BZ {(even though they are distinct EBRs)}.

Zak showed in Ref~\onlinecite{Zakphase} that because $\rho_G$ and $\rho'_G$ are induced from distinct Wyckoff positions, they have distinct Berry phases.
Namely, the holonomy of the Bloch wave function along a particular direction in momentum space gives the center of the Wannier function in the corresponding direction in real space.\cite{ksv}

Thus, the band representations $\rho_G$ and $\rho'_G$ are physically distinguishable, which motivates the need for Def~\ref{def:equiv}, a definition of equivalence that distinguishes them.
Physically, the reason these band representations are distinguishable is because there is no way to continuously move a single $s$ orbital from the $4a$ position at $(0,0,0)$ to the $4b$ position at $(0,0,1/2)$ while preserving the crystal symmetry because all of the {non-maximal} Wyckoff positions with sites immediately adjacent to the $4a$ position have a multiplicity greater than one; for example, a generic point on the $\hat{z}$ axis with coordinates $(0,0,z)$, where $z\neq 0,\frac{1}{2}$, belongs to a Wyckoff position with multiplicity two, which also contains $(0,0,-z)$, as depicted in Fig~\ref{fig:SG22horiz}. Thus, these two EBRs are not equivalent per our Def.~\ref{def:equiv}.

\section{Four-dimensional irreps of maximal site-symmetry groups}
\label{sec:irrepsdim4}

In this section, we prove that if $\rho$ is a four-dimensional irrep of a maximal site-symmetry group, $G_\mathbf{q}$,
then there does not exist a point, $\mathbf{q}_0$, whose site-symmetry group is an index two subgroup of $G_\mathbf{q}$.

There are only three point groups that have four-dimensional irreps: $O$, $T_d$ and $O_h$. 
$O$ and $T_d$ have one subgroup of index two, $T$, while $O_h$ has three subgroups of index two, $T_d$, $O$ and $T_h$.
All of the index two subgroups we have listed ($T, T_d, O$ and $T_h$) have a single fixed point, that is, given one of these subgroups, there is only one point which is left invariant by all of the operations in the subgroup.
We proved in Sec~\ref{sec:nopointfixed} that a site-symmetry group with a single fixed point must be maximal. 
It follows that if there exists a point, $\mathbf{q}_0$, which has $T, T_d, O$ or $T_h$ as its site-symmetry group, then these groups are a maximal subgroup of the space group.

Now consider a space group, $G$, such that there exists a site, $\mathbf{q}$, whose site-symmetry group, $G_\mathbf{q}$, is isomorphic to $O, T_d$ or $O_h$.
Further suppose that there exists a point, $\mathbf{q}_0$, whose site-symmetry group is given by one of the index two subgroups of $G_\mathbf{q}$; call this subgroup $G_0$.
We explained in the previous paragraph that $G_0$ must be a maximal subgroup of the space group.
However, this directly contradicts the definition of maximal (Def.~\ref{def:maximal}), since $G_0 \subset G_\mathbf{q} \subset G$ and $G_0 \neq G_\mathbf{q}$.
{  Thus, by contradiction, we have shown that if there exists a point $\mathbf{q}$, whose site-symmetry group, $G_\mathbf{q}$ is isomorphic to $O$, $T_d$ or $O_h$, then there does not exist a point $\mathbf{q}_0$ whose site-symmetry group is an index-two subgroup of $G_\mathbf{q}$.}
Since these are the only point groups with four-dimensional irreps, this completes the proof.

\section{Example of an exception in the honeycomb lattice}
\label{sec:exceptionexample}

Consider a band representation induced from the site $\mathbf{q} = (\mathbf{e}_1 - \mathbf{e}_2)/2$, which belongs to the Wyckoff position $3c$ of the honeycomb lattice, shown in Fig~\ref{fig:graphenebasisvectors}.
Utilizing the symmetry actions in Eq~(\ref{eq:grpaction2}), the site-symmetry group, $G_\mathbf{q}$, is generated by $\{ C_{2}| 1\bar{1} \}$ and $\{ m_{1\bar{1}}| 1\bar{1} \}$ and is isomorphic to $C_{2v}$, which has only one double-valued irrep, $\bar{\Gamma}_{5}$.

Since $G_\mathbf{q}$ is maximal (c.f. Def~\ref{def:maximal}), an irrep of $G_\mathbf{q}$ will induce an EBR, consisting of six bands, unless it is an exception, in the sense of Sec~\ref{sec:exceptions1}.
All exceptions for double-valued representations of three-dimensional space groups are listed in Table~\ref{table:dbr}; since the honeycomb lattice is two-dimensional, we consider its layered counterpart, $P6mm$ (SG 183), which does appear in Table~\ref{table:dbr}.
Hence, according to Table~\ref{table:dbr}, the band representation induced from the $\bar{\Gamma}_5$ irrep on $\mathbf{q}$ is equivalent (in the sense of Def~\ref{def:equiv}) to a composite band representation induced from either the $1a$ or $2b$ position, whose site-symmetry groups are isomorphic to $C_{6v}$ and $C_{3v}$, respectively.

We now prove explicitly that the band representation induced from $\mathbf{q}$ {in the absence of TR} is equivalent to a composite band representation induced from the $\mathbf{q}' = (0,0)$ position by constructing the homotopy described in Sec~\ref{sec:exceptions1}.

The site-symmetry group, $G_{\mathbf{q}'}$, of $\mathbf{q}'$ contains all symmetry operators that leave the origin invariant.
The only non-trivial element that leaves the origin invariant in $G_\mathbf{q}$ is $\{ m_{11}| \mathbf{0}\}$ (the product of $\{ C_{2}| 1\bar{1} \}$ and $\{ m_{1\bar{1}}| 1\bar{1} \}$).
Thus, the intersection of the two site-symmetry groups is given by, $G_0 \equiv G_\mathbf{q} \cap G_{\mathbf{q}'} = \{\{E| \mathbf{0} \},\{  m_{11}| \mathbf{0} \} \}$, which is isomorphic to $C_s$.
{$G_0$ is the site-symmetry group of the line connecting $\mathbf{q}$ and $\mathbf{q}'$.}
The character tables for $C_{6v}$, $C_{2v}$ and $C_s$ are shown in Tables~\ref{table:c6v}, \ref{table:c2v} and \ref{table:cs}.
Per the construction in Sec~\ref{sec:exceptions1}, we show that the representation of $G_\mathbf{q}$ induced from an irrep of $C_s$, ${\bar{\Gamma}_5} = \bar{\Gamma}_4 \uparrow G_\mathbf{q}$, is irreducible, while the representation of $G_{\mathbf{q}'}$ induced from an irrep of $C_s$, $\bar{\Gamma}_4 \uparrow G_{\mathbf{q}'}$, is reducible (we could also have induced representations from $\bar{\Gamma}_3$ and found the same result.)
First, since $C_s$ is an order 2 subgroup of $C_{2v}$, $\bar{\Gamma}_4$ will induce a two-dimensional representation of $C_{2v}$; since there is only one two-dimensional representation of $C_{2v}$, which is the irrep $\bar{\Gamma}_5$, we conclude that $\bar{\Gamma}_4 \uparrow C_{2v} = \bar{\Gamma}_5$.

Since $C_s$ is an order 6 subgroup of $C_{6v}$, $\bar{\Gamma}_4$ will induce a six-dimensional representation of $C_{6v}$, which is clearly reducible, as there is no $6D$ irrep of $C_{6v}$; however, this is not enough to uniquely determine the irreps into which it decomposes.
Instead, we compute the character of $C_{3}$ and $C_{6}$ in the induced representation $\bar{\Gamma}_4 \uparrow C_{6v}$, both of which are equal to zero using the Frobenius character formula\cite{Serre}. Specifically, we choose $C_{6}^n, n=0,...,5,$ to be the coset representatives of $C_{6v}/C_s$, and since $C_{6}^{-n}C_{3,6}C_{6}^n = C_{3,6}\notin C_s$ for all $n$, the character of $C_{3,6}$ in the induced representation is zero. Knowing that $\bar{\Gamma}_4 \uparrow C_{6v}$ is six-dimensional, this is enough to deduce from Table~\ref{table:c6v} that $\bar{\Gamma}_4 \uparrow C_{6v} = \bar{\Gamma}_7\oplus\bar{\Gamma}_8\oplus \bar{\Gamma}_9$ (in particular, $\bar{\Gamma}_8$ and $\bar{\Gamma}_9$ must appear equally in the decomposition because the character of $C_{6}$ is zero and $\bar{\Gamma}_7$ must also appear the same number of times {as $\bar{\Gamma}_8\oplus\bar{\Gamma}_9$} because the character of $C_{3}$ is zero).
Thus, we have shown that $\bar{\Gamma}_4 \uparrow C_{2v}$ is an irreducible representation of $C_{2v}$ while $\bar{\Gamma}_4 \uparrow C_{6v}$ is a reducible representation of $C_{6v}$.

We now notice that the line segment, $\alpha \mathbf{q}$, with $0<\alpha<1$, has $G_0$ as its site-symmetry group and, further, that the end points of the line are $(0,0)$ and $\mathbf{q}$.
Thus, the band representations induced from points on this line furnish a homotopy that smoothly connects the band representations induced from $\bar{\Gamma}_5$ on the $3c$ position and $\bar{\Gamma}_7\oplus\bar{\Gamma}_8\oplus \bar{\Gamma}_9$ on the $1a$ position.
Since the latter representation is composite (by Condition~\ref{cond:reducible}), the former is also composite.
This explicitly shows why $\bar{\Gamma}_5\uparrow G$ induced from the $3c$ position is composite, even though the $3c$ position is maximal.
However, if time reversal symmetry is imposed, $\bar{\Gamma}_5\uparrow G$ is a pEBR because the $1D$ homotopy does not obey time reversal symmetry in real space, as discussed in Sec~\ref{sec:exceptions2}. Thus, in a time reversal symmetric system, if the bands induced from the $3c$ position are gapped, then the gap is topological.

\section{Frobenius-Schur indicator: time reversal symmetry in real space}
\label{sec:Frobenius}

The reality of an irreducible representation, $\rho$, of a group, $G$,
is determined by computing the Frobenius-Schur indicator\cite{Fulton2004}: 
\begin{equation}
\Phi(\rho)=\frac{1}{|G|} \sum_{g\in G} \chi(g^2) = 
\begin{cases}
1 & \text{if $\rho$ is real} \\
0 & \text{if $\rho$ is complex}\\
-1 & \text{if $\rho$ is quaternionic}
\end{cases},
\label{eq:frobenius}
\end{equation}
where the sum is over all elements in $G$, including the identity.

If $\rho$ is real, then there exists an antiunitary time-reversal operator, $T_+$, {that commutes with all unitary symmetry operations and satisfies} $T_+^2=1$. If $\rho$ is a single-valued group representation, this means it is time-reversal invariant. On the other hand, if $\rho$ is a double-valued representation, this \emph{precludes} the possibility of finding a $T_-$ satisfying $T_-^2=-1$ (this follows from Schur's Lemma, c.~f.~Ref.~\onlinecite{Fulton2004}); in order to have a time-reversal invariant system we must double the representation to $\rho\oplus\rho$. We can then define $T_-=T_+\otimes(i\sigma_y)$.

The situation is reversed for $\Phi(\rho)=-1$. In this case, $\rho$ comes equipped with an antiunitary time-reversal operator $T_-$ satisfying $T_-^2=-1$. Thus, if $\rho$ is a double-valued group representation, it is time-reversal invariant. But, if $\rho$ is a single-valued group representation, we must double it to $\rho\oplus\rho$ in order to define $T_+=T_-\otimes(i\sigma_y)$ satisfying $T_+^2=1$.

Lastly, no time-reversal operation can be defined for complex representations, which satisfy $\Phi(\rho)=0$. For either single- or double-valued representations, the presence of time-reversal requires the representation to double to $\rho\oplus\rho^*$. This doubled representation is time-reversal invariant: for single group representations we then take $T=K\otimes\sigma_x$, while for double-valued group representations we take $T=K\otimes i\sigma_y$.

\bibliography{connectivity}
\end{document}